\documentclass{elsarticle}
\usepackage[utf8]{inputenc}
\usepackage[caption=false,font=footnotesize]{subfig}
\usepackage[nopdftrailerid=true]{pdfprivacy}
\usepackage{amssymb}
\usepackage{adjustbox}
\usepackage{tabularx}
\usepackage{makecell}
\usepackage{booktabs}
\usepackage{multirow}
\usepackage{datatool}
\usepackage[hyphens]{url}
\usepackage{hyperref}
\usepackage{bm}
\usepackage{algorithm}
\usepackage{algorithmic}
\usepackage{tikz}
\usetikzlibrary{positioning}

\newcommand{\DTLfetchsave}[5]{
	\edtlgetrowforvalue{#2}{\dtlcolumnindex{#2}{#3}}{#4}
	\dtlgetentryfromcurrentrow{\dtlcurrentvalue}{\dtlcolumnindex{#2}{#5}}
	\let#1\dtlcurrentvalue}

\pdftrailerid{}

\begin{document}
\begin{filecontents*}{python/artifacts/keys-values.csv}
,key,value
0,epochs-num,200
1,batch-size,2
2,num-slices-test,829
3,encoder-best,efficientnet-b4
4,encoder-worst,vgg19
5,lung-segmentation-None-mean,92.3
6,lung-segmentation-imagenet-mean,93.18
7,lesion-segmentation-a-None-mean,83.56
8,lesion-segmentation-a-imagenet-mean,85.47
9,lesion-segmentation-b-None-mean,85.44
10,lesion-segmentation-b-imagenet-mean,86.44
11,lung-segmentation-None-std,1.82
12,lung-segmentation-imagenet-std,1.3
13,lesion-segmentation-a-None-std,1.93
14,lesion-segmentation-a-imagenet-std,1.17
15,lesion-segmentation-b-None-std,1.08
16,lesion-segmentation-b-imagenet-std,1.04
17,lung-segmentation-None-max,95.08
18,lung-segmentation-imagenet-max,95.58
19,lesion-segmentation-a-None-max,86.83
20,lesion-segmentation-a-imagenet-max,87.56
21,lesion-segmentation-b-None-max,88.28
22,lesion-segmentation-b-imagenet-max,89.0
23,None-mean,87.1
24,None-std,88.36
25,imagenet-mean,4.1
26,imagenet-std,3.62
27,lung-segmentation-architecture-None-index-max,Unet
28,lung-segmentation-architecture-imagenet-index-max,resnet50
29,lesion-segmentation-a-architecture-None-index-max,Unet
30,lesion-segmentation-a-architecture-imagenet-index-max,resnet18
31,lesion-segmentation-b-architecture-None-index-max,Unet
32,lesion-segmentation-b-architecture-imagenet-index-max,dpn98
33,lung-segmentation-encoder-None-index-max,FPN
34,lung-segmentation-encoder-imagenet-index-max,xception
35,lesion-segmentation-a-encoder-None-index-max,Linknet
36,lesion-segmentation-a-encoder-imagenet-index-max,xception
37,lesion-segmentation-b-encoder-None-index-max,FPN
38,lesion-segmentation-b-encoder-imagenet-index-max,efficientnet-b4
39,lung-segmentation-Unet-None-mean,92.66
40,lung-segmentation-Unet-imagenet-mean,93.77
41,lung-segmentation-Linknet-None-mean,93.3
42,lung-segmentation-Linknet-imagenet-mean,93.96
43,lung-segmentation-FPN-None-mean,92.44
44,lung-segmentation-FPN-imagenet-mean,92.9
45,lung-segmentation-PSPNet-None-mean,90.81
46,lung-segmentation-PSPNet-imagenet-mean,92.09
47,lesion-segmentation-a-Unet-None-mean,84.33
48,lesion-segmentation-a-Unet-imagenet-mean,85.84
49,lesion-segmentation-a-Linknet-None-mean,84.25
50,lesion-segmentation-a-Linknet-imagenet-mean,86.14
51,lesion-segmentation-a-FPN-None-mean,83.25
52,lesion-segmentation-a-FPN-imagenet-mean,85.29
53,lesion-segmentation-a-PSPNet-None-mean,82.42
54,lesion-segmentation-a-PSPNet-imagenet-mean,84.61
55,lesion-segmentation-b-Unet-None-mean,85.35
56,lesion-segmentation-b-Unet-imagenet-mean,86.86
57,lesion-segmentation-b-Linknet-None-mean,86.05
58,lesion-segmentation-b-Linknet-imagenet-mean,86.5
59,lesion-segmentation-b-FPN-None-mean,85.46
60,lesion-segmentation-b-FPN-imagenet-mean,86.52
61,lesion-segmentation-b-PSPNet-None-mean,84.91
62,lesion-segmentation-b-PSPNet-imagenet-mean,85.86
63,lung-segmentation-Unet-None-std,2.04
64,lung-segmentation-Unet-imagenet-std,0.95
65,lung-segmentation-Linknet-None-std,1.19
66,lung-segmentation-Linknet-imagenet-std,0.96
67,lung-segmentation-FPN-None-std,1.29
68,lung-segmentation-FPN-imagenet-std,1.25
69,lung-segmentation-PSPNet-None-std,1.63
70,lung-segmentation-PSPNet-imagenet-std,1.06
71,lesion-segmentation-a-Unet-None-std,1.31
72,lesion-segmentation-a-Unet-imagenet-std,0.64
73,lesion-segmentation-a-Linknet-None-std,1.0
74,lesion-segmentation-a-Linknet-imagenet-std,0.79
75,lesion-segmentation-a-FPN-None-std,2.31
76,lesion-segmentation-a-FPN-imagenet-std,1.17
77,lesion-segmentation-a-PSPNet-None-std,2.08
78,lesion-segmentation-a-PSPNet-imagenet-std,1.33
79,lesion-segmentation-b-Unet-None-std,1.03
80,lesion-segmentation-b-Unet-imagenet-std,0.83
81,lesion-segmentation-b-Linknet-None-std,0.96
82,lesion-segmentation-b-Linknet-imagenet-std,0.82
83,lesion-segmentation-b-FPN-None-std,0.83
84,lesion-segmentation-b-FPN-imagenet-std,1.1
85,lesion-segmentation-b-PSPNet-None-std,1.15
86,lesion-segmentation-b-PSPNet-imagenet-std,1.12
\end{filecontents*}
\DTLloaddb{keys-values}{python/artifacts/keys-values.csv}

\title{Comprehensive Comparison of Deep Learning Models for Lung and COVID-19 Lesion Segmentation in CT scans}
\author[1]{Paschalis Bizopoulos}
\author[1]{Nicholas Vretos}
\author[1]{Petros Daras}

\address[1]{Visual Computing Lab of the Information Technologies Institute, Centre for Research and Technology Hellas, Thessaloniki, Greece e-mail: pbizop@gmail.com,\{vretos,daras\}@iti.gr}

\begin{abstract}
	Deep Learning (DL) methods are extensively used for medical image segmentation.
	However the field's reliability is hindered by the lack of a common base of reference for accuracy/performance evaluation and that previous research uses different datasets for evaluation.
	In this paper, an extensive comparison of DL models for lung and COVID-19 lesion segmentation in Computerized Tomography (CT) scans is presented, which can also be used as a benchmark for testing medical image segmentation models.
	Four DL architectures (Unet, Linknet, FPN, PSPNet) are combined with $25$ randomly initialized and pretrained encoders (variations of VGG, DenseNet, ResNet, ResNext, DPN, MobileNet, Xception, Inception-v4, EfficientNet), to construct $200$ tested models.
	Three experimental setups are conducted for lung segmentation, lesion segmentation and lesion segmentation using the original lung masks.
	A COVID-19 dataset with $100$ CT scan images is used for training/validation and a different dataset consisting of $\DTLfetch{keys-values}{key}{num-slices-test}{value}$ images from $9$ CT scan volumes for testing.
	Multiple findings are provided including the best architecture-encoder models for each experiment as well as mean Dice results for each experiment, architecture and encoder independently.
	The source code and $600$ pretrained models for the three experiments are provided, suitable for fine-tuning in experimental setups without GPU capabilities.
\end{abstract}


\maketitle

\section{Introduction}\label{sec:introduction}
Coronavirus Disease 2019 (COVID-19) has emerged in December of 2019 and was declared as a pandemic in March of 2020~\cite{world2020coronavirus}.
The Severe Acute Respiratory Syndrome Coronavirus 2 (SARS-CoV-2) has certain properties that makes it highly infectious, thus, turning ineffective government policy measures such as social distancing and increasing the need for fast and accurate diagnosis of the disease.
A well established, high resolution, imaging procedure that targets lungs and depicts rich pathological information is Computerized Tomography (CT) scan.
More specifically, for a COVID-19 patient, CT scan images show bilateral patchy shadows or ground glass opacity on the infected region~\cite{wang2020clinical}, which are not always visible in common X-Ray scans~\cite{ng2020imaging}.
Another method that has been used for COVID-19 diagnosis is the so-called Reverse-Transcription Polymerase Chain Reaction (RT-PCR), which, however, has been found to have lower sensitivity compared to CT~\cite{ai2020correlation} scan and is more time consuming.

Medical experts often need to examine a large number of CT scan images, which is an error prone and time consuming process.
To that aim, automatic segmentation methods are being proposed that segment regions-of-interest (ROIs) of different size and shape such as lungs, nodules and lesions, taking advantage of the CT scan resolution.
These methods facilitate medical experts in diagnosing by focusing on the ROIs instead of the whole image.
Methods for automatic segmentation in the lung area from the literature include the use of morphological operations~\cite{hu2001automatic}, active contours~\cite{keshani2013lung} and fuzzy clustering~\cite{manikandan2016lung}.

Feature engineering methods however, were surpassed by end-to-end learning such as Deep Learning (DL)~\cite{lecun2015deep}, which were successfully applied in medical image segmentation tasks~\cite{minaee2020image}.
More specifically, applications of DL methods in medical image segmentation primarily target lungs~\cite{skourt2018lung, jin2020ai}, pathological lungs~\cite{harrison2017progressive}, infections~\cite{fan2020inf, chen2020residual, wang2020noise}, lungs and infections~\cite{yan2020covid}, lungs and COVID-19 lesions~\cite{shan2020lung}.
The majority of them uses encoder-decoder architectures such as Unet~\cite{ronneberger2015u} and its variations.

A major issue in the field of lung/lesion segmentation (and medical image segmentation in general) is the use of different datasets for evaluating newly proposed models.
Moreover, there is lack of benchmark baseline models that could play the role of reference for evaluating the accuracy and the performance of proposed models.
Benchmarks for COVID-19 in CT scan images were provided in the literature, such as Ma et al.~\cite{ma2020towards} that has a limited number of cases and He et al.~\cite{he2020benchmarking} that test $20$ models for lung segmentation of COVID-19 patients.
Other comparison studies on similar tasks such as lung nodule segmentation were proposed in~\cite{kalpathy2016comparison}, which compare three non-learnable algorithms, where each one is created by a different research group.
Comparison studies of deep learning image segmentation tasks has also been conducted on non-medical images such as coral reef images~\cite{king2018comparison} where the authors test four models as well as aerial city images~\cite{liu2018comparison} where the authors test $12$ different models.
No previous work, to the best of our knowledge, includes a comprehensive quantified comparison of $600$ DL models on the task of image segmentation.

In this paper, four of the most widely used DL image segmentation architectures are explored, namely Unet~\cite{ronneberger2015u}, Linknet~\cite{chaurasia2017linknet}, Feature Pyramid Network (FPN)~\cite{lin2017feature} and Pyramid Scene Parsing Network (PSPNet)~\cite{zhao2017pyramid} combined with $25$ encoders for lung and COVID-19 lesion segmentation in CT scan images.
The contribution of this paper in the field of medical image segmentation can be summarized as follows:
\begin{itemize}
	\item derivation of best architecture-encoder combinations for the three experiments that are conducted (lung, lesion and lesion with lung masks experiments),
	\item quantitative comparison of architectures,
	\item quantitative comparison of encoders,
	\item quantitative comparison of lesion segmentation with and without masks (in this case lungs) as a preprocessing step,
	\item quantitative comparison of random and ImageNet initialization,
	\item open source implementation\footnote{\url{https://github.com/pbizopoulos/comprehensive-comparison-of-deep-learning-models-for-lung-and-covid-19-lesion-segmentation-in-ct}},
	\item release of $600$ pretrained models of all experimental setups for use by external researchers.
\end{itemize}

The rest of the paper is organized as follows: a detailed description of the models and their components (architecture, encoder) is provided in Section~\ref{sec:methods}, the datasets used are presented in Section~\ref{sec:datasets}, the experimental setup used to evaluate the models is shown in Section~\ref{sec:experimentalsetup} and the results are demonstrated in Section~\ref{sec:results}. Finally the findings in relation with findings in the previous literature are shown in Section~\ref{sec:discussion} and the final remarks are concluded in Section~\ref{sec:conclusions}.

\section{Methods}\label{sec:methods}
In this Section the problem of segmentation is formalized and the architectures and encoders that are used in this study are presented.

\subsection{Formalization}
\begin{figure}[!t]
	\centering
	\begin{tikzpicture}[]
		\node[] (input) {\includegraphics[scale=0.3]{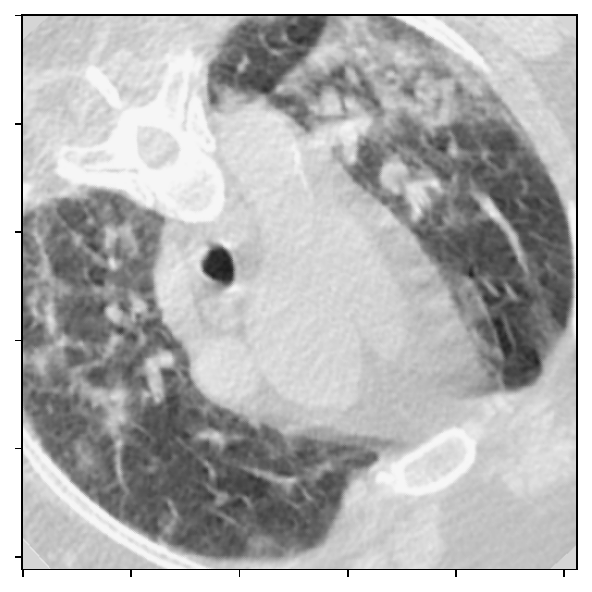}};
		\node[draw] (architecture) [right=0.5cm of input, minimum height=2.8cm, minimum width=1.2cm]{\hspace{-2em}\rotatebox{90}{architecture}};
		\node[draw] (encoder) [right=1cm of input, minimum height=2cm]{\rotatebox{90}{encoder}};
		\node[] (output) [right=0.5cm of encoder]{\includegraphics[scale=0.3]{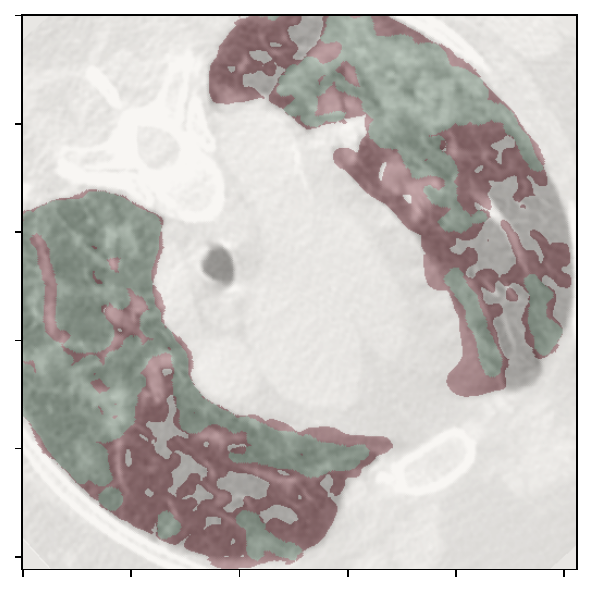}};
		\node[draw, below=0.5cm of output, circle] (loss){$\mathcal{L}$};
		\node[] (mask) [below=0.5cm of loss]{\includegraphics[scale=0.3]{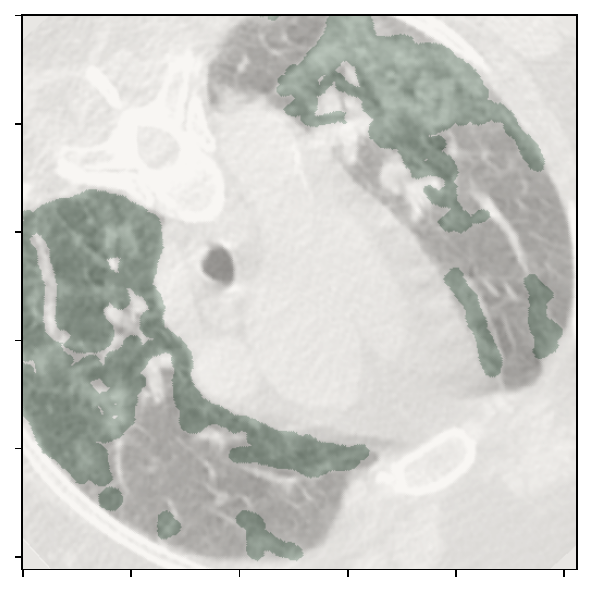}};
		\node[] at (0, 1){$\bm{X}$};
		\node[] at (5.2, 1){$\hat{\bm{Y}}$};
		\node[] at (5.2, -4){$\bm{Y}$};
		\draw[->] (input) -- node{} (architecture);
		\draw[->] (architecture) -- node{} (output);
		\draw[->] (output) -- node{} (loss);
		\draw[->] (mask) -- node{} (loss);
		\draw[->] (loss) -| node[shift={(0, -0.2)}]{backpropagation} (architecture);
	\end{tikzpicture}
	\caption{High level overview of a lesion segmentation model with an architecture consisting of an encoder trained using a single training augmented image.
	$\bm{X}$ is the input, $\bm{Y}$ is the target mask, $\hat{\bm{Y}}$ is the predicted mask and $\mathcal{L}$ is the loss function, which in this case is the Dice loss.
	Green and red pixels at $\hat{\bm{Y}}$ depict correctly and falsely classified pixels, while green pixels at $\bm{Y}$ depict the pixels of the target mask.
	Arrows denote the flow of the feed-forward and backpropagation pass.
	$\bm{X}$ is passed to the architecture consisting of a specific encoder and $\hat{\bm{Y}}$ is calculated.
	Then $\hat{\bm{Y}}$ and $\bm{Y}$ are used to calculate the loss, which is then used to backpropagate the error to the weights.
	This procedure is repeated for a number of times using more training examples till $\mathcal{L}$ converges.}\label{fig:highleveloverview}
\end{figure}
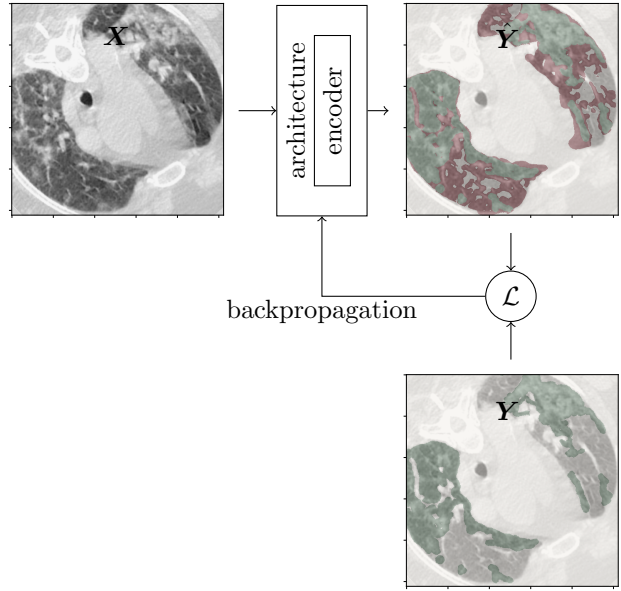

Let $\mathcal{D}$ be a dataset containing images $\bm{X} \in \mathbb{R}^{n_r, n_c}$ and $\bm{Y} \in \{0, 1\}^{n_r, n_c}$ the corresponding target mask (in our case $n_r=512$, $n_c=512$).
Let $m_{ex, ar, en, ew}$ be a DL model for segmentation where $ex$ denotes the specific experiment, $ar$ the architecture, $en$ the encoder and $ew$ the encoder weights.
The `encoder' is defined as the part of the model that performs the feature extraction.
The model $m_{ex, ar, en, ew}$ is trained on a dataset $\mathcal{D}_{train}\subset\mathcal{D}$ consisting of $\bm{X}_{train} \in \mathbb{R}^{n_r, n_c}$ and $\bm{Y}_{train} \in \{0, 1\}^{n_r, n_c}$.
Moreover a validation dataset is defined as $\mathcal{D}_{val}\subset\mathcal{D}$ where $\mathcal{D}_{val}\cap\mathcal{D}_{train}=\emptyset$ consisting of $\bm{X}_{val} \in \mathbb{R}^{n_r, n_c}$ and $\bm{Y}_{val} \in \{0, 1\}^{n_r, n_c}$.
Therefore, the objective of the experiments conducted in this study can be designed as, finding an optimal point in the parameter space of $m_{ex, ar, en, ew}$ during training such that when presented with an input from $\mathcal{D}$ such as $\bm{X}_{val}$, its prediction $\hat{\bm{Y}}_{val} \in [0, 1]^{n_r, n_c}$ is as near as possible with the target $\bm{Y}_{val}$.
This is implemented by selecting the model that performs the minimum validation error out of all epochs.
Subsequently, the selected models are tested on the generalization ability on an unseen $\mathcal{D}_{test}$ with $\mathcal{D}_{test}\cap\mathcal{D}=\emptyset$.
A high level overview of the training of the models can be seen in Fig.~\ref{fig:highleveloverview}.

\subsection{Architectures}\label{sec:architectures}
Four architectures are used as the basis of the models to be tested:
\begin{itemize}
	\item Unet~\cite{ronneberger2015u}
	\item Linknet~\cite{chaurasia2017linknet}
	\item Feature Pyramid Network (FPN)~\cite{lin2017feature}
	\item Pyramid Scene Parsing Network (PSPNet)~\cite{zhao2017pyramid}
\end{itemize}

Unet~\cite{ronneberger2015u} combines an encoder that scales down the features to a lower dimensional bottleneck and a decoder that scales them up to original dimensions.
It also uses skip connections that were proven to improve image segmentation results~\cite{drozdzal2016importance}.
Linknet~\cite{chaurasia2017linknet} is similar to Unet with the difference of using residual~\cite{he2016deep} instead of convolutional blocks in its encoder and decoder networks.
Feature Pyramid Network (FPN)~\cite{lin2017feature} is also similar to Unet with the difference of applying a $1\times1$ convolution layer and adding the features instead of copying and appending them as done in the Unet architecture.
The Pyramid Scene Parsing Network (PSPNet)~\cite{zhao2017pyramid} exploits a pyramid pooling module to aggregate the image global context information with an auxiliary loss~\cite{hu2019comparison}.

\subsection{Encoders}\label{sec:encoders}
The following encoders are used along with their variations denoted in the parenthesis:
\begin{itemize}
	\item VGG~\cite{simonyan2014very} (11, 13, 19)
	\item DenseNet~\cite{huang2017densely} (121, 161, 169, 201)
	\item ResNet~\cite{he2016deep} (18, 34, 50, 101, 152)
	\item ResNext~\cite{xie2017aggregated}
	\item Dual Path Networks (DPN)~\cite{chen2017dual} (68, 98)
	\item MobileNet~\cite{howard2017mobilenets}
	\item Xception~\cite{chollet2017xception}
	\item Inception-v4~\cite{szegedy2017inception}
	\item EfficientNet~\cite{tan2019efficientnet} (b0, b1, b3, b4, b5, b6)
\end{itemize}

VGG~\cite{simonyan2014very} is named after the Visual Geometry Group that proposed it and took the second place during the ImageNet Competition in 2014~\cite{deng2009imagenet}.
It was one of the first models that demonstrated the importance of depth in DL and it is preferred for tasks such as feature extraction due to its simple repeating structure.
On the other hand, ResNet's~\cite{he2016deep} (abbreviation of Residual Networks) contributions allowed training deep networks by using layers that learn residual functions with reference to layer inputs, while DenseNet~\cite{huang2017densely} uses connections between each layer and every other layer in a feed-forward fashion.
Moreover ResNext~\cite{xie2017aggregated} consists of a stack of residual blocks, which are subject to two rules.
The first one is that layers that output spatial maps with the same size, share hyper-parameters and the second is that when a spatial map is downsampled by two, the width of the blocks is multiplied by two.
Dual Path Networks (DPN)~\cite{chen2017dual} proposed as a network that combines feature re-usage and feature exploration that ResNet and DenseNet do respectively, while MobileNet~\cite{howard2017mobilenets} constructed to fill the need of training and inferencing on devices with low computational capabilities such as embedded device and mobile phones.
Xception~\cite{chollet2017xception} is a variation of Inception Network~\cite{szegedy2015going} in which the inception modules have been replaced with depthwise convolutions followed by a pointwise convolution.
Finally Inception-v4~\cite{szegedy2017inception} combines previous inception architectures with residual connections achieving state-of-the-art performance on the ImageNet, while EfficientNet~\cite{tan2019efficientnet} is an improvement of MobileNet where the compound scaling module was proposed as an efficient way to uniformly scale depth, width and resolution.

\section{Datasets}\label{sec:datasets}
Two public COVID-19 CT scan datasets with lung and lesion masks were used.
The first dataset\footnote{\url{http://medicalsegmentation.com/covid19/}} consists of $100$ CT axial scans from $<40$ patients with $512\times 512$ size and corresponding lung masks from~\cite{hofmanninger2020automatic} and lesion masks labeled with four classes (none, ground-glass, consolidation, pleural effusion).
The original dataset, without the annotations, was selected from the Italian Society of Medical and Interventional Radiology\footnote{\url{https://www.sirm.org/category/senza-categoria/covid-19/}}.
The second dataset\footnote{\url{https://radiopaedia.org/articles/covid-19-3}} consists of $\DTLfetch{keys-values}{key}{num-slices-test}{value}$ images from $9$ CT\@ scan volumes (a set of CT scan images acquired from the same patient at the same moment) with corresponding target masks.
373 out of $\DTLfetch{keys-values}{key}{num-slices-test}{value}$ were annotated as positive and segmented by a radiologist by the same group as the first dataset.
Raw data from both datasets contain samples in Hounsfield units~\cite{schneider1996calibration}.

Regarding preprocessing, first the positive classes of the pixels of the images in the first dataset are merged into one, converting the problem to a binary segmentation problem.
The CT scan images from the second dataset are resized to $512\times 512$, and both datasets are normalized with $\mu=-500$ and $\sigma=500$.
We use $80$ scans from the first dataset for training the models, $20$ scans for validation and all scans from the second dataset for the testing of the models.

\begin{figure*}[!t]
	\rotatebox[origin=l]{90}{\hspace{1em}\scriptsize Normalized frequencies}\subfloat[Lung segmentation]{\includegraphics[width=0.33\textwidth]{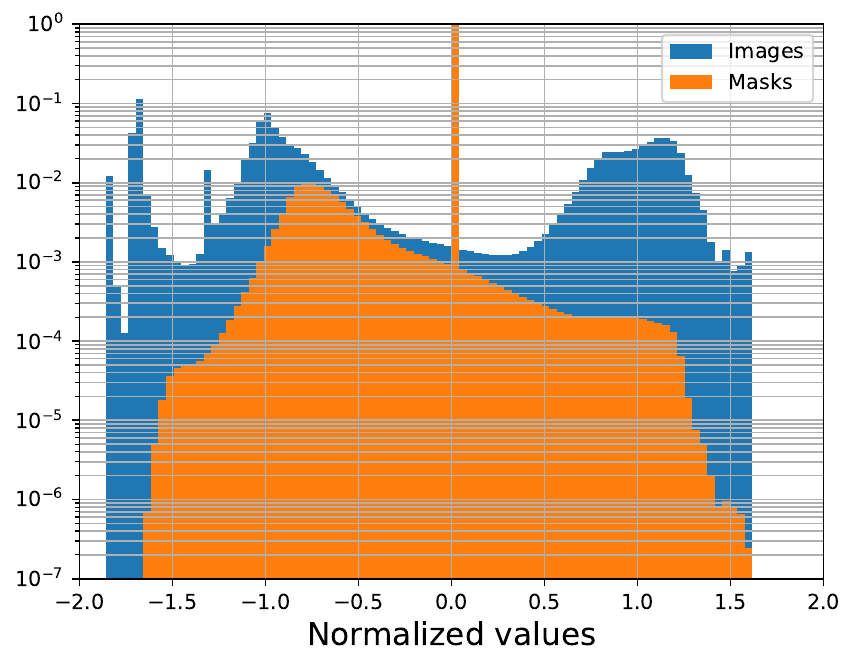}}
	\subfloat[Lesion segmentation A]{\includegraphics[width=0.33\textwidth]{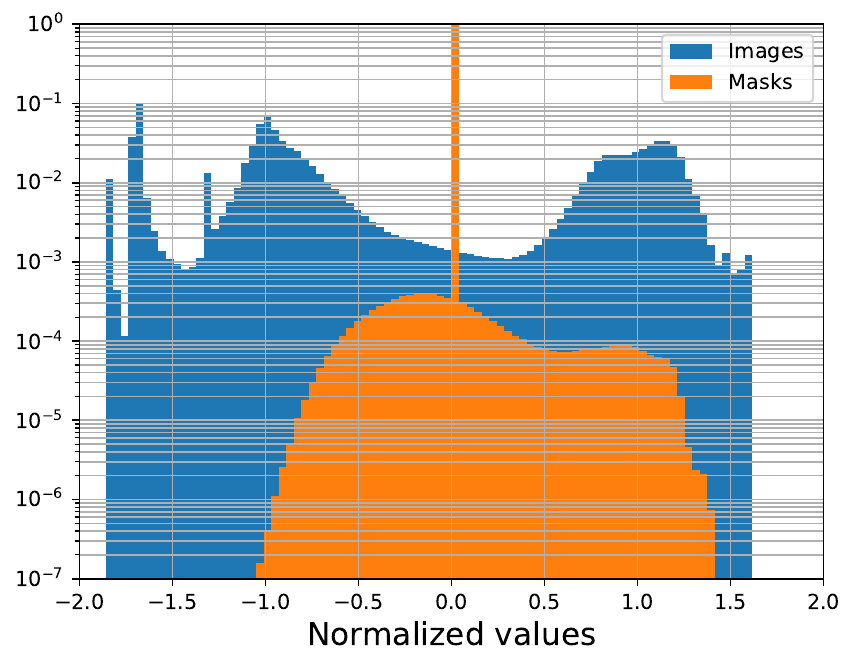}}
	\subfloat[Lesion segmentation B]{\includegraphics[width=0.33\textwidth]{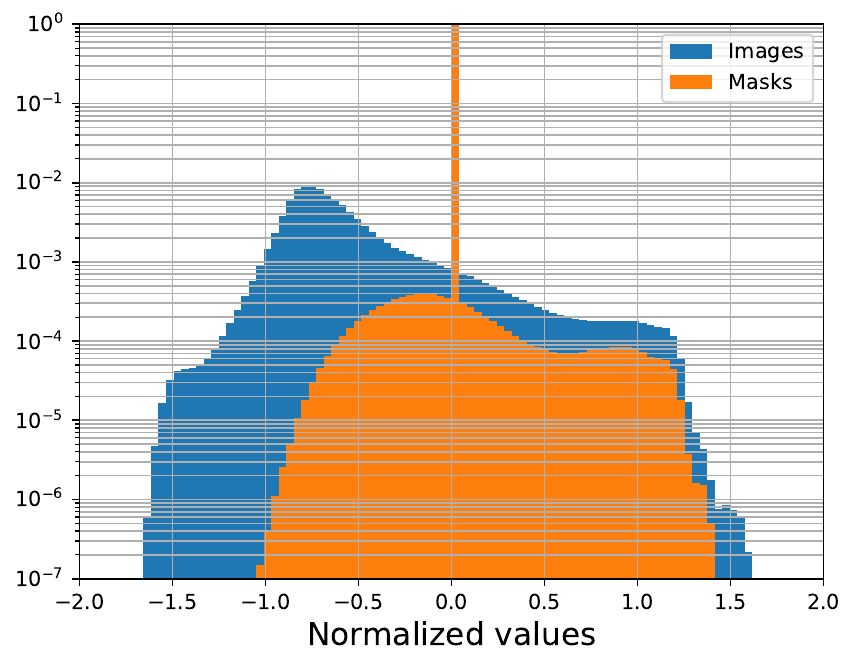}}
	\caption{Histogram plots of the normalized pixel intensities for the images (blue) and target masks (orange) for each experimental setup.
	The vertical axis depicts the normalized frequencies of each pixel intensity in the logarithmic scale.
	The outlier bar at zero is because most of the pixels in the target masks are near zero.}\label{fig:hist}
\end{figure*}

In Fig.~\ref{fig:hist} the histograms of the pixel intensities of all the CT scan images and the target masks in the test dataset after normalization, for each of the three experiments, are depicted.
The considerable overlap between the histograms makes the use of thresholding models in this kind of problem unsuitable, thus justifying the use of learning models such as DL\@.

\section{Experimental setup}\label{sec:experimentalsetup}
In this Section the experimental setup is presented.
In total three experiments are conducted:
\begin{itemize}
	\item lung segmentation
	\item lesion segmentation (referred to as `lesion segmentation A')
	\item lesion segmentation with lung masks (referred to as `lesion segmentation B')
\end{itemize}

The choice of these experiments covers balanced (lung segmentation), unbalanced (lesion segmentation A) and unbalanced with preprocessing (lesion segmentation B) image segmentation tasks and the findings could apply in image segmentation tasks with non-medical images.
Each of the performed experiments uses a different target mask $\bm{Y}$, where for the `lesion segmentation B' the corresponding lung masks is also applied in the input image $\bm{X}$.

Each model is constructed using a unique combination of the four architectures described in Subsection~\ref{sec:architectures} and the $25$ encoders referenced in Subsection~\ref{sec:encoders}.
The selection of architectures and encoders was based on the restriction of the GPU memory of our graphics card, combined with the value of batch size.
Then, for each model we also test the randomly initialized and its ImageNet pretrained version.

The default values for every hyperparameter of the models were used (as seen in Table.~\ref{table:architecturehyperparameters}), to avoid favoring models that were proposed after being evaluated in a controlled experimental setup.
The activation function for all architectures was sigmoid that squashes the output in the range of $[0, 1]$.

\begin{table}[]
	\centering
	\caption{Architecture hyperparameters}\label{table:architecturehyperparameters}
	\begin{tabular}{lccc}
		\toprule
		\textbf{Architecture} & \textbf{\makecell{Encoder\\ depth}}          & \textbf{Batch Norm}                & \textbf{Various}                                                                                             \\
		Unet    & 5 & Yes (decoder) & \makecell{decoder channel sizes =\\ (256, 128, 64, 32, 16)}                                 \\
		\midrule
		Linknet & 5 & Yes (decoder) & -                                                                              \\
		\midrule
		FPN     & 5 & No            & \makecell{pyramid channels=256,\\ segment channels=128,\\ merge policy=add,\\ dropout=0.2} \\
		\midrule
		PSPNet  & 3 & Yes (encoder) & \makecell{output channels=512,\\ dropout=0.2}                                              \\
		\bottomrule
	\end{tabular}
\end{table}

For all experiments and in each epoch during training, data augmentation is applied on the images from the training dataset:
\begin{itemize}
	\item horizontal/vertical flip each with probability 50\%,
	\item rotation with an angle chosen from a uniform distribution with range $[-180^{\circ}, 180^{\circ}]$ and
	\item scale within a range of $[0.5, 1.5]$ with zero padding
\end{itemize}

During training the Soft Dice Loss is used to calculate the error of the model on the training dataset as:
\begin{equation}
	Soft Dice Loss = 1 - 2\frac{\sum\limits^{n_r}_i\sum\limits^{n_c}_j\bm{Y}_{ij}\hat{\bm{Y}}_{ij}}{\sum\limits^{n_r}_i\sum\limits^{n_c}_j\bm{Y}_{ij}^2 + \sum\limits^{n_r}_i\sum\limits^{n_c}_j\hat{\bm{Y}}_{ij}^2 + \epsilon},
\end{equation}

where $\bm{Y}_{ij}$, $\hat{\bm{Y}}_{ij}$ are the pixel intensities at the $i^{th}$ column, $j^{th}$ row of the target mask and predicted mask, respectively (which applies for $\bm{Y}_{train}$, $\bm{Y}_{val}$ and $\bm{Y}_{test}$) and $\epsilon=10^{-5}$.
The model selection is done using Soft Dice Loss in each epoch during training on the validation dataset.
During testing, the predicted mask $\hat{\bm{Y}}_{test}$ is binarized with a threshold value of $0.5$ allowing us to use hard metrics for testing the models:
\begin{equation}
	Sensitivity = \frac{TP}{TP + FN + \epsilon}
\end{equation}
\begin{equation}
	Specificity = \frac{TN}{TN + FP + \epsilon}
\end{equation}
\begin{equation}
	Dice = \frac{2TP}{2TP + FP + FN + \epsilon}
\end{equation}

where $TP$, $TN$, $FP$ and $FN$ are the true positive, true negative, false positive and false negative of $\hat{\bm{Y}}$ w.r.t $\bm{Y}$, respectively and $\epsilon=10^{-5}$ to prevent division with zero.
When $TP+FP=0$ the model has correctly identified that the input does not have any positive pixel and in that case all metrics are set to $1$.

We train a total of $600$ different models for the three experiments each one for $\DTLfetch{keys-values}{key}{epochs-num}{value}$ epochs with a batch size of $\DTLfetch{keys-values}{key}{batch-size}{value}$, which was the maximum possible considering the GPU memory restriction.
We use the optimizer~\cite{kingma2014adam} with the default values of learning rate $0.001$, $\beta_1=0.9$, $\beta_2=0.999$, $\epsilon=10^{-8}$, without weight decay.
Pytorch~\cite{paszke2019pytorch} and the `Segmentation Models Pytorch' library~\cite{yakubovskiy2019segmentation} were used for implementing the experiments, a GeForce RTX 2080 Ti Graphics Card with 11Gb RAM from NVIDIA and an Intel Core i9-9900K CPU @3.60GHz, on a Linux-based operating system for training the models for two weeks.
A pseudo-code implementation of the experimental setup is shown in Algorithm~\ref{alg:experimentalsetup}.

\begin{algorithm}[H]
	\caption{Experimental setup}\label{alg:experimentalsetup}
	\begin{algorithmic}[1]
		\renewcommand{\algorithmicrequire}{\textbf{Input:}}
		\renewcommand{\algorithmicensure}{\textbf{Output:}}
		\REQUIRE{$epochs$}
		\ENSURE{$metrics$}
		\\ \textit{Hyperparameters:  $\lambda$, $batches$}
		\FOR{$ex$ = 1 to $n_{experiments}$}
		\FOR{$ar$ = 1 to $n_{architectures}$}
		\FOR{$en$ = 1 to $n_{encoders}$}
		\FOR{$ew$ = 1 to $n_{encoder\ weights}$}
		\FOR{$ep$ = 1 to $epochs$}
		\FOR{$b$ = 1 to $batches$}
		\STATE{$\bm{Y}_{train}, \bm{X}_{train} \sim \mathcal{D}_{train}$}
		\STATE{$\bm{Y}_{train}, \bm{X}_{train} \leftarrow Augm(\bm{Y}_{train}, \bm{X}_{train})$}
		\STATE{$\hat{\bm{Y}}_{train} \leftarrow m_{ex, ar, en, ew}(\bm{X}_{train})$}
		\STATE{$\mathcal{L}_{train} \leftarrow DiceLoss(\hat{\bm{Y}}_{train}, \bm{Y}_{train})$}
		\STATE{$\nabla\mathcal{L}_{train} = \left( \frac{\partial\mathcal{L}}{\partial\bm{w}^{(1)}},\ldots\frac{\partial\mathcal{L}}{\partial\bm{w}^{(q)}}\right)$}
		\STATE{$\Delta\bm{w}^{(i)} \leftarrow -\lambda\frac{\partial\mathcal{L}}{\partial\bm{w}^{(i)}}$}
		\ENDFOR{}
		\FOR{$b$ = 1 to $batches$}
		\STATE{$\bm{Y}_{val}, \bm{X}_{val} \sim \mathcal{D}_{val}$}
		\STATE{$\hat{\bm{Y}}_{val} \leftarrow m_{ex, ar, en, ew}(\bm{X}_{val})$}
		\STATE{$\mathcal{L}_{val} \leftarrow DiceLoss(\hat{\bm{Y}}_{val}, \bm{Y}_{val})$}
		\ENDFOR{}
		\IF{$\mathcal{L}_{val} < \mathcal{L}_{val}^{best}$}
		\STATE{$m_{ex, ar, en, ew}^{best} \leftarrow m_{ex, ar, en, ew}$}
		\ENDIF{}
		\ENDFOR{}
		\STATE{$\hat{\bm{Y}}_{test} \leftarrow m_{ex, ar, en, ew}^{best}(\bm{X}_{test})$}
		\STATE{$metrics_{ex, ar, en, ew} = metrics(\hat{\bm{Y}}_{test}, \bm{Y}_{test})$}
		\ENDFOR{}
		\ENDFOR{}
		\ENDFOR{}
		\ENDFOR{}
		\RETURN{$metrics$}
	\end{algorithmic}
\end{algorithm}

\section{Results}\label{sec:results}
In this Section the results of the three experimental setups are demonstrated, along with several comparisons between experiments, architectures, encoders and weight initialization schemes.

\subsection{Overall}
\begin{figure}[!t]
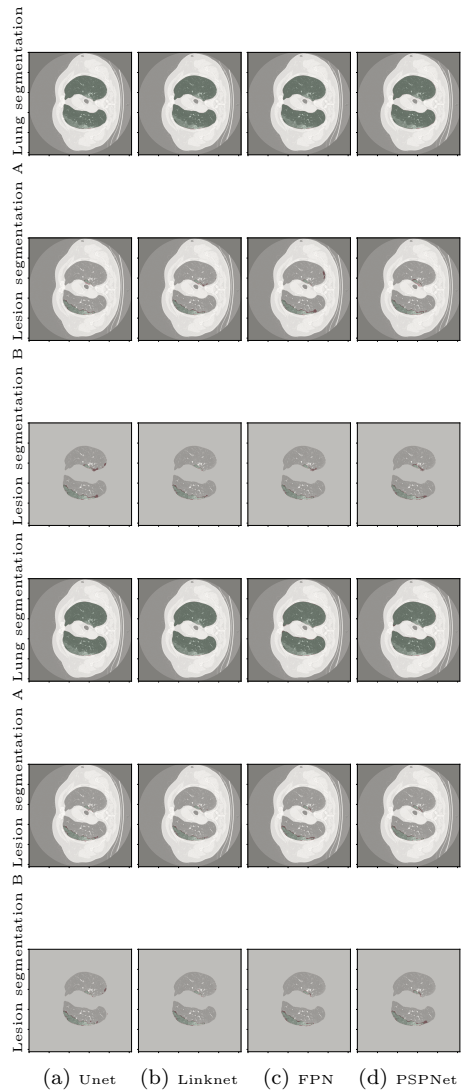

	\centering
	\rotatebox[origin=l]{90}{\tiny Lung segmentation}\subfloat{\includegraphics[width=0.12\textwidth]{python/artifacts/lung-segmentation-Unet-\best-masked-image}}
	\subfloat{\includegraphics[width=0.12\textwidth]{python/artifacts/lung-segmentation-Linknet-\best-masked-image}}
	\subfloat{\includegraphics[width=0.12\textwidth]{python/artifacts/lung-segmentation-FPN-\best-masked-image}}
	\subfloat{\includegraphics[width=0.12\textwidth]{python/artifacts/lung-segmentation-PSPNet-\best-masked-image}}
	\\
	\rotatebox[origin=l]{90}{\tiny Lesion segmentation A}\subfloat{\includegraphics[width=0.12\textwidth]{python/artifacts/lesion-segmentation-a-Unet-\best-masked-image}}
	\subfloat{\includegraphics[width=0.12\textwidth]{python/artifacts/lesion-segmentation-a-Linknet-\best-masked-image}}
	\subfloat{\includegraphics[width=0.12\textwidth]{python/artifacts/lesion-segmentation-a-FPN-\best-masked-image}}
	\subfloat{\includegraphics[width=0.12\textwidth]{python/artifacts/lesion-segmentation-a-PSPNet-\best-masked-image}}
	\\
	\rotatebox[origin=l]{90}{\tiny Lesion segmentation B}\subfloat{\includegraphics[width=0.12\textwidth]{python/artifacts/lesion-segmentation-b-Unet-\best-masked-image}}
	\subfloat{\includegraphics[width=0.12\textwidth]{python/artifacts/lesion-segmentation-b-Linknet-\best-masked-image}}
	\subfloat{\includegraphics[width=0.12\textwidth]{python/artifacts/lesion-segmentation-b-FPN-\best-masked-image}}
	\subfloat{\includegraphics[width=0.12\textwidth]{python/artifacts/lesion-segmentation-b-PSPNet-\best-masked-image}}
	\\
	\rotatebox[origin=l]{90}{\tiny Lung segmentation}\subfloat{\includegraphics[width=0.12\textwidth]{python/artifacts/lung-segmentation-Unet-\best-masked-image}}
	\subfloat{\includegraphics[width=0.12\textwidth]{python/artifacts/lung-segmentation-Linknet-\worst-masked-image}}
	\subfloat{\includegraphics[width=0.12\textwidth]{python/artifacts/lung-segmentation-FPN-\worst-masked-image}}
	\subfloat{\includegraphics[width=0.12\textwidth]{python/artifacts/lung-segmentation-PSPNet-\worst-masked-image}}
	\\
	\rotatebox[origin=l]{90}{\tiny Lesion segmentation A}\subfloat{\includegraphics[width=0.12\textwidth]{python/artifacts/lesion-segmentation-a-Unet-\worst-masked-image}}
	\subfloat{\includegraphics[width=0.12\textwidth]{python/artifacts/lesion-segmentation-a-Linknet-\worst-masked-image}}
	\subfloat{\includegraphics[width=0.12\textwidth]{python/artifacts/lesion-segmentation-a-FPN-\worst-masked-image}}
	\subfloat{\includegraphics[width=0.12\textwidth]{python/artifacts/lesion-segmentation-a-PSPNet-\worst-masked-image}}
	\\
	\setcounter{subfigure}{0}
	\rotatebox[origin=l]{90}{\tiny Lesion segmentation B}\subfloat[{\tiny Unet}]{\includegraphics[width=0.12\textwidth]{python/artifacts/lesion-segmentation-b-Unet-\worst-masked-image}}
	\subfloat[{\tiny Linknet}]{\includegraphics[width=0.12\textwidth]{python/artifacts/lesion-segmentation-b-Linknet-\worst-masked-image}}
	\subfloat[{\tiny FPN}]{\includegraphics[width=0.12\textwidth]{python/artifacts/lesion-segmentation-b-FPN-\worst-masked-image}}
	\subfloat[{\tiny PSPNet}]{\includegraphics[width=0.12\textwidth]{python/artifacts/lesion-segmentation-b-PSPNet-\worst-masked-image}}
	\caption{Predicted masks on a CT scan image from the test data for $24$ out of $600$ of the models.
	The three top columns correspond to predicted masks generated from a model with the best encoder and the bottom three for the worst.
	Rows correspond to each of the four architectures.
	Green and red depict correctly and falsely classified pixels respectively.}\label{fig:predictedmasks}
\end{figure}

\begin{figure}[!t]
	\centering
	\rotatebox[origin=l]{90}{\hspace{2em}\scriptsize Dice (\%)}\subfloat[Lung segmentation]{\includegraphics[width=0.24\textwidth]{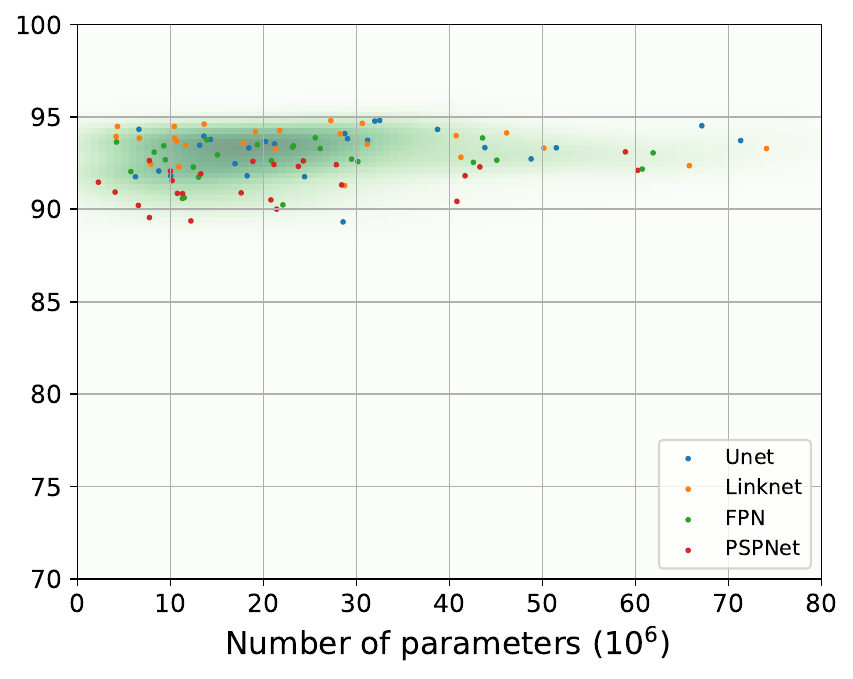}}
	\subfloat[Lesion segmentation A]{\includegraphics[width=0.24\textwidth]{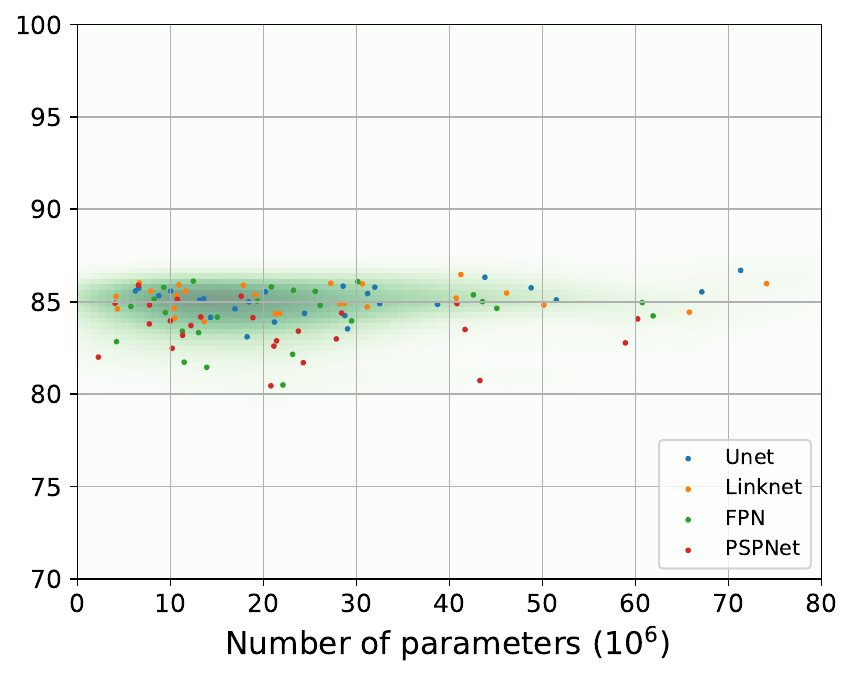}}
	\\
	\rotatebox[origin=l]{90}{\hspace{2em}\scriptsize Dice (\%)}\subfloat[Lesion segmentation B]{\includegraphics[width=0.24\textwidth]{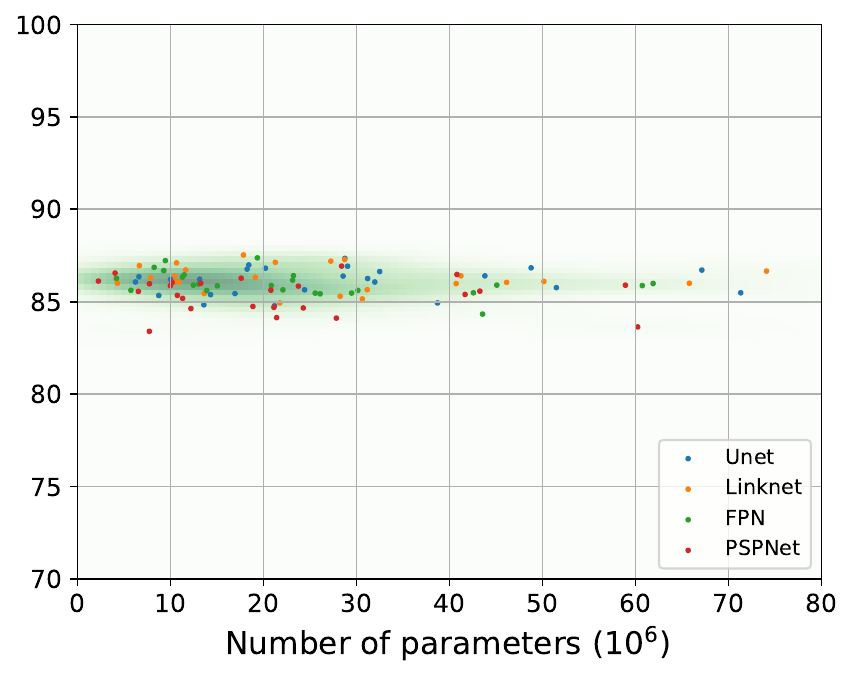}}
	\subfloat[Lesion segmentation B-A]{\includegraphics[width=0.24\textwidth]{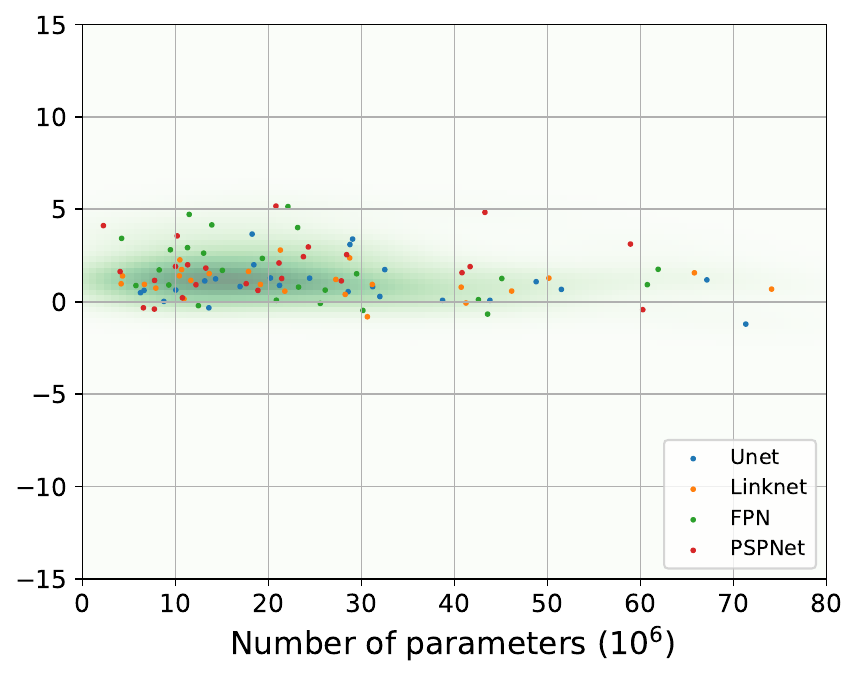}}
	\caption{Dice vs.\ number of parameters for the three experimental setups and the difference between lesion segmentation B and A.
	The green regions depict the kernel density estimate for all points using Gaussian kernels.
	The method used to calculate the estimator bandwidth was Scott and the points are assumed to be equally weighted.}\label{fig:dicevsnumparameters}
\end{figure}

In Table.~\ref{table:metrics} the resulted metrics of all experiments are presented.
The best combination of architecture-encoder for each combination of encoder weight initialization, experimental setup and metric are showin in bold.
The mean Dice results for each experiment are $\DTLfetch{keys-values}{key}{lung-segmentation-imagenet-mean}{value}\%\pm\DTLfetch{keys-values}{key}{lung-segmentation-imagenet-std}{value}\%$ for lung segmentation, $\DTLfetch{keys-values}{key}{lesion-segmentation-a-imagenet-mean}{value}\%\pm\DTLfetch{keys-values}{key}{lesion-segmentation-a-imagenet-std}{value}\%$, for `lesion segmentation A' and $\DTLfetch{keys-values}{key}{lesion-segmentation-b-imagenet-mean}{value}\%\pm\DTLfetch{keys-values}{key}{lesion-segmentation-b-imagenet-std}{value}\%$, for `lesion segmentation B'.
The best performing models w.r.t.\ Dice for each experiment were the \DTLfetch{keys-values}{key}{lung-segmentation-architecture-imagenet-index-max}{value}-\DTLfetch{keys-values}{key}{lung-segmentation-encoder-imagenet-index-max}{value} ($\DTLfetch{keys-values}{key}{lung-segmentation-imagenet-max}{value}\%$) for lung segmentation, \DTLfetch{keys-values}{key}{lesion-segmentation-a-architecture-imagenet-index-max}{value}-\DTLfetch{keys-values}{key}{lesion-segmentation-a-encoder-imagenet-index-max}{value} ($\DTLfetch{keys-values}{key}{lesion-segmentation-a-imagenet-max}{value}\%$) for lesion segmentation A and \DTLfetch{keys-values}{key}{lesion-segmentation-b-architecture-imagenet-index-max}{value}-\DTLfetch{keys-values}{key}{lesion-segmentation-b-encoder-imagenet-index-max}{value} ($\DTLfetch{keys-values}{key}{lesion-segmentation-b-imagenet-max}{value}\%$) for lesion segmentation B.
In Fig.~\ref{fig:predictedmasks} the predicted masks for $24$ out of the $600$ models are depicted, demonstrating the difference in segmentation quality between the best (\DTLfetch{keys-values}{key}{encoder-best}{value}) and the worst (\DTLfetch{keys-values}{key}{encoder-worst}{value}) performing encoder of the models for each architecture, with randomly initialized weights.
In Fig.~\ref{fig:dicevsnumparameters} the Dice vs.\ the number of parameters is plotted, demonstrating that there is positive correlation, suggesting that segmentation generally improves when using higher number of parameters.
However, this is not a significant positive correlation.
It is worth noting that the best model is not the one with the largest number of parameters.

\DTLfetchsave{\best}{keys-values}{key}{encoder-best}{value}
\DTLfetchsave{\worst}{keys-values}{key}{encoder-worst}{value}

\begin{table*}[!t]
	\centering
	\caption{Metrics}\label{table:metrics}
	\begin{adjustbox}{width=\textwidth}
		\begin{tabular}{clrrrrrrrrrrrrrrrrrrrrr}
\toprule
       &     & \multicolumn{9}{c}{None} & \multicolumn{9}{c}{imagenet} & \multicolumn{3}{c}{Performance} \\
       &     & \multicolumn{3}{c}{Lung segmentation} & \multicolumn{3}{c}{Lesion segmentation A} & \multicolumn{3}{c}{Lesion segmentation B} & \multicolumn{3}{c}{Lung segmentation} & \multicolumn{3}{c}{Lesion segmentation A} & \multicolumn{3}{c}{Lesion segmentation B} & \multicolumn{3}{c}{related} \\
       &     &              Sens &       Spec &       Dice &                  Sens &       Spec &       Dice &                  Sens &       Spec &       Dice &              Sens &       Spec &       Dice &                  Sens &       Spec &       Dice &                  Sens &       Spec &       Dice &     Pars(M) &  Train(s) &    Val(s) \\
\textbf{Architecture} & \textbf{Encoder} &                   &            &            &                       &            &            &                       &            &            &                   &            &            &                       &            &            &                       &            &            &             &           &           \\
\midrule
\multirow{27}{*}{\textbf{Unet}} & \textbf{vgg11} &             89.03 &      99.82 &      91.00 &                 79.47 &      99.87 &      81.49 &                 81.60 &      99.95 &      85.85 &             91.08 &      99.88 &      92.63 &                 81.79 &      99.92 &      84.71 &                 84.07 &      99.94 &      87.67 &       18.25 &      0.32 &      0.06 \\
       & \textbf{vgg13} &             92.80 &      99.77 &      92.41 &                 83.35 &      99.87 &      84.55 &                 81.57 & \textbf{99.96} &      86.02 &             92.52 &      99.89 &      94.24 &                 82.73 &      99.91 &      85.43 &                 84.46 &      99.94 &      87.95 &       18.44 &      0.37 &      0.07 \\
       & \textbf{vgg19} &             92.73 &      99.83 &      94.07 &                 80.36 &      99.87 &      81.78 &                 83.20 &      99.94 &      87.22 &             91.96 &      99.87 &      93.57 &                 83.84 &      99.90 &      85.27 &                 83.06 &      99.93 &      86.62 &       29.06 &      0.44 &      0.08 \\
       & \textbf{resnet18} &             89.68 &      99.85 &      91.98 &                 80.69 &      99.86 &      82.26 &                 79.79 &      99.93 &      83.92 &             94.61 &      99.87 & \textbf{95.58} &                 81.97 &      99.95 &      86.04 &                 83.76 &      99.92 &      86.85 &       14.32 &      0.50 &      0.09 \\
       & \textbf{resnet34} &             87.51 &      99.78 &      89.75 &                 79.75 &      99.89 &      82.98 &                 82.17 &      99.91 &      85.52 &             91.76 &      99.91 &      93.78 &                 81.56 &      99.95 &      85.75 &                 81.92 &      99.93 &      85.76 &       24.43 &      0.71 &      0.12 \\
       & \textbf{resnet50} &             94.21 &      99.83 & \textbf{95.08} &                 81.85 &      99.88 &      84.10 &                 83.24 &      99.92 &      86.63 &             94.33 &      99.85 &      94.54 &                 81.38 &      99.95 &      85.69 &                 83.33 &      99.92 &      86.63 &       32.51 &      0.94 &      0.16 \\
       & \textbf{resnet101} &             93.35 &      99.79 &      91.60 &                 83.08 &      99.90 &      85.25 &                 80.85 &      99.93 &      85.01 &             93.68 &      99.87 &      95.06 &                 81.30 &      99.93 &      84.94 &                 82.79 &      99.94 &      86.51 &       51.51 &      1.62 &      0.28 \\
       & \textbf{resnet152} &        \textbf{94.31} &      99.81 &      94.90 &                 82.15 &      99.93 &      85.55 &                 83.61 &      99.91 &      86.58 &             92.26 &      99.87 &      94.15 &                 82.13 &      99.93 &      85.51 &                 83.16 &      99.94 &      86.84 &       67.15 &      2.27 &      0.39 \\
       & \textbf{densenet121} &             93.92 &      99.78 &      93.66 &                 81.95 &      99.89 &      83.64 &                 79.45 &      99.94 &      83.98 &             92.77 &      99.86 &      94.26 &                 82.24 & \textbf{99.96} &      86.69 &                 81.97 &      99.93 &      85.69 &       13.60 &      2.32 &      0.42 \\
       & \textbf{densenet161} &             92.67 &      99.88 &      94.62 &                 81.80 &      99.91 &      84.21 &                 82.04 &      99.92 &      85.25 &             92.11 &      99.86 &      94.02 &                 81.16 &      99.95 &      85.51 &                 80.79 &      99.92 &      84.62 &       38.73 &      3.25 &      0.59 \\
       & \textbf{densenet169} &             91.26 &      99.84 &      93.08 &                 80.44 &      99.87 &      82.11 &                 79.48 &      99.93 &      83.52 &             91.96 &      99.91 &      93.98 &                 81.56 &      99.94 &      85.68 &                 82.24 &      99.94 &      86.04 &       21.20 &      3.40 &      0.61 \\
       & \textbf{densenet201} &             82.66 &      99.70 &      85.61 &                 83.15 &      99.92 &      85.68 &                 81.46 &      99.94 &      85.59 &             91.07 &      99.89 &      93.02 &                 81.83 &      99.95 &      86.00 &                 84.17 &      99.91 &      87.18 &       28.57 &      4.06 &      0.78 \\
       & \textbf{resnext5032x4d} &             93.49 &      99.88 &      94.57 &                 82.56 &      99.93 &      85.43 &                 82.77 &      99.94 &      86.15 &             93.26 &      99.90 &      94.98 &                 81.64 & \textbf{99.96} &      86.14 &                 81.58 &      99.96 &      85.99 &       31.99 &      1.43 &      0.29 \\
       & \textbf{dpn68} &             88.70 &      99.88 &      91.54 &                 80.32 &      99.92 &      83.96 &                 80.79 &      99.93 &      84.44 &             91.38 &      99.89 &      93.41 &                 80.94 &      99.95 &      85.27 &                 83.29 &      99.92 &      86.44 &       16.95 &      2.09 &      0.42 \\
       & \textbf{dpn98} &             92.66 &      99.84 &      93.89 &            \textbf{85.15} &      99.91 & \textbf{86.83} &                 79.93 &      99.94 &      84.44 &             91.48 &      99.90 &      93.55 &                 83.77 &      99.93 &      86.56 &                 83.25 &      99.93 &      86.52 &       71.33 &      3.13 &      0.65 \\
       & \textbf{mobilenetv2} &             93.52 &      99.87 &      94.95 &                 82.06 &      99.91 &      84.93 &                 81.12 &      99.95 &      85.51 &             91.82 &      99.91 &      93.71 &                 82.84 &      99.94 &      86.55 &                 83.13 &      99.95 &      87.19 &        6.63 &      0.87 &      0.15 \\
       & \textbf{xception} &             92.94 &      99.85 &      94.07 &                 80.58 &      99.92 &      83.61 &                 84.07 &      99.93 &      87.06 &             93.06 &      99.89 &      94.13 &                 80.18 & \textbf{99.96} &      84.88 &                 83.80 &      99.94 &      87.63 &       28.77 &      0.94 &      0.16 \\
       & \textbf{inceptionv4} &             92.08 &      99.82 &      93.28 &                 83.82 &      99.87 &      84.88 &                 82.21 &      99.93 &      85.99 &             91.82 &      99.86 &      92.17 &                 82.45 & \textbf{99.96} &      86.62 &                 83.84 &      99.95 &      87.68 &       48.79 &      2.41 &      0.42 \\
       & \textbf{efficientnet-b0} &             87.72 &      99.86 &      89.99 &                 81.76 &      99.92 &      84.73 &                 81.52 &      99.94 &      85.81 &             92.20 &      99.90 &      93.53 &                 83.44 &      99.93 &      86.43 &                 82.06 &      99.94 &      86.33 &        6.25 &      1.42 &      0.25 \\
       & \textbf{efficientnet-b1} &             90.53 &      99.89 &      92.80 &                 83.33 &      99.88 &      85.36 &                 79.01 &      99.95 &      83.80 &             89.73 &      99.91 &      91.35 &                 81.19 &      99.94 &      85.28 &                 82.29 &      99.95 &      86.88 &        8.76 &      1.91 &      0.33 \\
       & \textbf{efficientnet-b2} &             89.17 &      99.88 &      91.66 &                 82.16 &      99.90 &      84.63 &                 80.93 &      99.92 &      84.94 &             89.90 &      99.91 &      91.98 &                 82.39 &      99.95 &      86.53 &                 83.62 &      99.94 &      87.49 &       10.05 &      1.92 &      0.34 \\
       & \textbf{efficientnet-b3} &             91.14 &      99.89 &      93.19 &                 82.18 &      99.91 &      85.10 &                 79.42 &      99.94 &      83.99 &             92.47 &      99.90 &      93.74 &                 80.06 & \textbf{99.96} &      85.08 &                 84.86 &      99.94 &      88.44 &       13.16 &      2.09 &      0.38 \\
       & \textbf{efficientnet-b4} &             90.89 &      99.91 &      93.13 &                 80.28 &      99.94 &      84.65 &                 82.06 &      99.94 &      86.31 &             92.84 &      99.89 &      94.20 &                 82.00 & \textbf{99.96} &      86.41 &                 83.05 &      99.95 &      87.32 &       20.22 &      2.56 &      0.46 \\
       & \textbf{efficientnet-b5} &             91.29 &      99.90 &      93.41 &                 80.60 &      99.95 &      84.86 &                 80.30 &      99.93 &      84.75 &             92.42 &      99.91 &      94.06 &                 81.84 &      99.95 &      86.02 &                 84.41 &      99.94 &      87.76 &       31.22 &      3.13 &      0.55 \\
       & \textbf{efficientnet-b6} &             90.38 &      99.79 &      92.19 &                 82.83 &      99.91 &      85.56 &                 81.16 &      99.94 &      85.39 &             92.74 &      99.91 &      94.48 &                 84.06 &      99.93 &      87.09 &                 83.65 &      99.93 &      87.41 &       43.82 &      3.54 &      0.62 \\
       & \textbf{Mean} &             91.15 &      99.84 &      92.66 &                 81.83 &      99.90 &      84.33 &                 81.35 &      99.93 &      85.35 &             92.21 &      99.89 &      93.77 &                 82.01 &      99.94 &      85.84 &                 83.14 &      99.94 &      86.86 &       27.83 &      1.91 &      0.35 \\
       & \textbf{Std} &              2.62 &       0.05 &       2.04 &                  1.38 &       0.02 &       1.31 &                  1.37 &       0.01 &       1.03 &              1.12 &       0.02 &       0.95 &                  1.01 &       0.02 &       0.64 &                  0.97 &       0.01 &       0.83 &       17.29 &      1.08 &      0.20 \\
\cline{1-23}
\multirow{27}{*}{\textbf{Linknet}} & \textbf{vgg11} &             92.04 &      99.87 &      93.85 &                 79.99 &      99.90 &      82.66 &                 82.97 &      99.94 &      86.38 &             92.25 &      99.86 &      93.85 &                 81.67 &      99.94 &      85.57 &                 82.74 &      99.94 &      86.38 &       10.48 &      0.38 &      0.07 \\
       & \textbf{vgg13} &             92.67 &      99.82 &      92.82 &                 83.14 &      99.91 &      85.43 &                 82.11 &      99.95 &      86.38 &             93.26 &      99.89 &      94.57 &                 81.95 &      99.93 &      85.29 &                 84.88 &      99.92 &      87.82 &       10.67 &      0.40 &      0.07 \\
       & \textbf{vgg19} &             90.31 &      99.83 &      92.62 &                 82.18 &      99.89 &      83.50 &                 83.83 &      99.94 &      87.62 &             92.97 &      99.87 &      93.96 &                 81.43 &      99.93 &      85.17 &                 83.31 &      99.92 &      86.64 &       21.29 &      0.46 &      0.08 \\
       & \textbf{resnet18} &             90.61 &      99.87 &      92.94 &                 81.31 &      99.91 &      84.45 &                 83.46 &      99.94 &      86.69 &             92.35 &      99.89 &      93.97 &                 83.30 &      99.94 &      86.69 &                 83.19 &      99.92 &      86.75 &       11.66 &      0.55 &      0.09 \\
       & \textbf{resnet34} &             93.06 &      99.85 &      94.56 &                 81.15 &      99.89 &      83.41 &                 80.92 &      99.93 &      84.91 &             92.46 &      99.86 &      93.97 &                 81.10 & \textbf{99.96} &      85.32 &                 80.81 &      99.95 &      84.96 &       21.77 &      0.80 &      0.14 \\
       & \textbf{resnet50} &             92.70 &      99.74 &      93.16 &                 81.38 &      99.91 &      84.34 &                 81.76 &      99.94 &      85.77 &             94.01 &      99.83 &      93.88 &                 80.54 & \textbf{99.96} &      85.09 &                 81.28 &      99.95 &      85.53 &       31.17 &      1.01 &      0.18 \\
       & \textbf{resnet101} &             93.56 &      99.69 &      92.26 &                 81.98 &      99.92 &      85.06 &                 80.86 &      99.94 &      85.09 &             93.37 &      99.82 &      94.36 &                 80.95 &      99.94 &      84.58 &                 83.55 &      99.94 &      87.10 &       50.16 &      1.69 &      0.29 \\
       & \textbf{resnet152} &             89.45 &      99.66 &      89.62 &                 80.45 &      99.91 &      83.19 &                 83.33 &      99.91 &      86.75 &             93.89 &      99.84 &      95.10 &                 82.84 &      99.93 &      85.68 &                 81.31 &      99.94 &      85.25 &       65.81 &      2.34 &      0.40 \\
       & \textbf{densenet121} &             92.34 &      99.80 &      93.76 &                 79.31 &      99.90 &      82.50 &                 82.16 &      99.93 &      85.78 &             93.57 &      99.90 &      95.21 &                 83.28 &      99.94 &      86.79 &                 82.36 &      99.95 &      86.32 &       10.42 &      2.44 &      0.44 \\
       & \textbf{densenet161} &             93.22 &      99.82 &      94.43 &                 81.02 &      99.92 &      84.51 &                 81.74 &      99.92 &      85.54 &             91.51 &      99.91 &      93.54 &                 81.99 &      99.94 &      85.88 &                 83.78 &      99.91 &      86.43 &       40.73 &      3.30 &      0.62 \\
       & \textbf{densenet169} &             93.10 &      99.81 &      94.16 &                 80.49 &      99.94 &      84.53 &                 82.98 &      99.94 &      86.38 &             92.49 &      99.89 &      94.25 &                 82.29 &      99.95 &      86.23 &                 82.07 &      99.95 &      86.25 &       19.14 &      3.54 &      0.64 \\
       & \textbf{densenet201} &             92.71 &      99.79 &      93.58 &                 81.58 &      99.89 &      83.92 &                 79.85 &      99.95 &      84.28 &             93.51 &      99.83 &      94.58 &                 81.84 &      99.95 &      85.89 &                 83.11 &      99.92 &      86.31 &       28.26 &      4.22 &      0.81 \\
       & \textbf{resnext5032x4d} &             93.42 &      99.86 &      94.40 &                 82.49 &      99.91 &      84.95 &                 79.44 &      99.95 &      84.11 &             93.20 &      99.91 &      94.90 &                 83.36 &      99.95 &      86.98 &                 82.39 &      99.94 &      86.18 &       30.64 &      1.56 &      0.31 \\
       & \textbf{dpn68} &             93.76 &      99.87 &      94.74 &                 78.95 &      99.92 &      82.90 &                 81.43 &      99.95 &      85.46 &             92.61 &      99.91 &      94.49 &                 80.81 &      99.95 &      84.96 &                 81.58 &      99.93 &      85.43 &       13.64 &      2.21 &      0.44 \\
       & \textbf{dpn98} &             92.85 &      99.85 &      93.75 &                 83.36 &      99.92 &      85.87 &                 83.80 &      99.93 &      87.16 &             92.42 &      99.83 &      92.82 &                 82.04 &      99.95 &      86.08 &                 82.65 &      99.93 &      86.16 &       74.11 &      3.30 &      0.68 \\
       & \textbf{mobilenetv2} &             92.16 &      99.88 &      94.00 &                 79.51 &      99.90 &      82.12 &                 82.68 &      99.92 &      86.05 &        \textbf{94.88} &      99.86 &      94.97 &                 83.89 &      99.93 &      87.10 &                 81.03 &      99.96 &      85.96 &        4.32 &      0.96 &      0.17 \\
       & \textbf{xception} &             92.88 &      99.87 &      94.37 &                 81.66 &      99.93 &      84.91 &            \textbf{85.05} &      99.94 & \textbf{88.28} &             93.97 &      99.88 &      95.24 &                 83.43 &      99.95 &      87.07 &                 81.54 &      99.96 &      86.11 &       27.26 &      1.02 &      0.18 \\
       & \textbf{inceptionv4} &             94.13 &      99.76 &      94.20 &                 82.62 &      99.87 &      83.78 &                 81.83 &      99.90 &      84.95 &             93.99 &      99.85 &      94.07 &                 83.32 &      99.94 &      87.16 &                 82.80 &      99.96 &      87.15 &       46.16 &      2.45 &      0.43 \\
       & \textbf{efficientnet-b0} &             92.46 &      99.81 &      93.69 &                 81.84 &      99.90 &      84.27 &                 80.85 &      99.95 &      85.64 &             92.34 &      99.91 &      94.19 &                 82.41 &      99.94 &      86.29 &                 82.45 &      99.94 &      86.86 &        4.17 &      1.43 &      0.27 \\
       & \textbf{efficientnet-b1} &             92.62 &      99.86 &      94.26 &                 82.68 &      99.90 &      84.98 &                 82.62 &      99.92 &      86.40 &             92.75 &      99.87 &      93.44 &                 83.20 &      99.95 &      87.04 &                 83.39 &      99.94 &      87.49 &        6.67 &      2.01 &      0.35 \\
       & \textbf{efficientnet-b2} &             90.37 &      99.87 &      92.31 &                 81.29 &      99.92 &      84.59 &                 81.48 &      99.94 &      85.69 &             91.06 &      99.89 &      92.59 &                 83.46 &      99.93 &      86.55 &                 82.49 &      99.95 &      86.93 &        7.89 &      2.01 &      0.35 \\
       & \textbf{efficientnet-b3} &             89.71 &      99.83 &      91.37 &                 81.85 &      99.93 &      84.95 &                 82.07 &      99.94 &      86.32 &             92.30 &      99.86 &      93.22 &                 83.08 &      99.94 &      86.87 &                 80.62 &      99.96 &      85.80 &       10.93 &      2.19 &      0.40 \\
       & \textbf{efficientnet-b4} &             92.48 &      99.81 &      93.77 &                 80.81 &      99.95 &      85.05 &                 82.49 &      99.94 &      86.39 &             91.35 & \textbf{99.92} &      93.44 &                 82.84 &      99.94 &      86.73 &            \textbf{85.31} &      99.94 &      88.66 &       17.86 &      2.74 &      0.47 \\
       & \textbf{efficientnet-b5} &             90.03 &      99.87 &      91.89 &                 79.73 & \textbf{99.96} &      84.47 &                 83.60 &      99.93 &      87.28 &             88.25 &      99.90 &      90.68 &                 80.71 & \textbf{99.96} &      85.36 &                 83.31 &      99.94 &      87.28 &       28.74 &      3.22 &      0.56 \\
       & \textbf{efficientnet-b6} &             89.81 & \textbf{99.93} &      91.97 &                 82.97 &      99.92 &      85.90 &                 81.31 & \textbf{99.96} &      86.00 &             92.12 &      99.88 &      93.66 &                 83.49 &      99.95 &      87.04 &                 82.17 &      99.96 &      86.80 &       41.25 &      3.64 &      0.64 \\
       & \textbf{Mean} &             92.10 &      99.82 &      93.30 &                 81.35 &      99.91 &      84.25 &                 82.18 &      99.94 &      86.05 &             92.67 &      99.87 &      93.96 &                 82.37 &      99.94 &      86.14 &                 82.57 &      99.94 &      86.50 &       25.41 &      2.00 &      0.36 \\
       & \textbf{Std} &              1.38 &       0.06 &       1.19 &                  1.20 &       0.02 &       1.00 &                  1.28 &       0.01 &       0.96 &              1.27 &       0.03 &       0.96 &                  1.01 &       0.01 &       0.79 &                  1.15 &       0.01 &       0.82 &       18.44 &      1.11 &      0.21 \\
\cline{1-23}
\multirow{27}{*}{\textbf{FPN}} & \textbf{vgg11} &             88.12 &      99.84 &      90.24 &                 79.18 &      99.91 &      82.83 &                 81.51 &      99.94 &      85.82 &             88.98 &      99.82 &      90.94 &                 79.12 &      99.95 &      83.99 &                 82.68 &      99.93 &      86.84 &       11.30 &      0.27 &      0.06 \\
       & \textbf{vgg13} &             87.91 &      99.87 &      89.94 &                 76.46 &      99.92 &      80.27 &                 81.52 &      99.93 &      85.73 &             89.36 &      99.89 &      91.31 &                 78.34 &      99.95 &      83.18 &                 82.80 &      99.95 &      87.17 &       11.49 &      0.30 &      0.06 \\
       & \textbf{vgg19} &             89.12 &      99.77 &      90.75 &                 73.33 &      99.90 &      77.93 &                 81.45 &      99.86 &      84.81 &             88.09 &      99.78 &      89.72 &                 79.25 &      99.92 &      83.06 &                 82.23 &      99.93 &      86.48 &       22.11 &      0.38 &      0.07 \\
       & \textbf{resnet18} &             89.24 &      99.78 &      90.93 &                 76.83 &      99.90 &      80.81 &                 81.92 &      99.95 &      86.10 &             91.04 &      99.83 &      92.56 &                 81.46 &      99.95 &      85.85 &                 81.18 &      99.96 &      85.81 &       13.04 &      0.48 &      0.09 \\
       & \textbf{resnet34} &             92.77 &      99.75 &      94.07 &                 75.13 &      99.90 &      79.33 &                 81.26 &      99.94 &      85.42 &             90.86 &      99.85 &      92.67 &                 80.19 & \textbf{99.96} &      84.98 &                 83.22 &      99.95 &      86.92 &       23.15 &      0.73 &      0.12 \\
       & \textbf{resnet50} &             92.98 &      99.76 &      93.96 &                 80.12 &      99.94 &      84.56 &                 79.95 &      99.93 &      84.42 &             90.27 &      99.85 &      92.62 &                 80.24 &      99.94 &      85.03 &                 81.99 &      99.95 &      86.43 &       26.11 &      0.91 &      0.16 \\
       & \textbf{resnet101} &             92.05 &      99.81 &      93.89 &                 79.66 &      99.94 &      84.37 &                 82.51 &      99.92 &      86.17 &             89.60 &      99.82 &      91.43 &                 80.13 & \textbf{99.96} &      84.91 &                 81.25 &      99.95 &      85.63 &       45.10 &      1.61 &      0.28 \\
       & \textbf{resnet152} &             90.00 &      99.68 &      91.22 &                 79.78 &      99.92 &      83.61 &                 80.74 &      99.93 &      85.22 &             90.98 &      99.87 &      93.14 &                 82.77 &      99.94 &      86.29 &                 82.30 &      99.94 &      86.53 &       60.75 &      2.27 &      0.39 \\
       & \textbf{densenet121} &             92.42 &      99.77 &      93.77 &                 81.72 &      99.92 &      84.97 &                 81.84 &      99.95 &      86.40 &             91.84 &      99.81 &      93.11 &                 82.93 &      99.94 &      86.58 &                 83.35 &      99.93 &      86.97 &        9.29 &      2.29 &      0.42 \\
       & \textbf{densenet161} &             91.24 &      99.85 &      93.35 &                 79.75 &      99.91 &      83.83 &                 79.96 &      99.93 &      84.35 &             90.42 &      99.87 &      92.08 &                 78.91 & \textbf{99.96} &      84.09 &                 82.80 &      99.94 &      86.59 &       29.49 &      3.09 &      0.58 \\
       & \textbf{densenet169} &             91.17 &      99.80 &      92.96 &                 80.67 &      99.92 &      83.63 &                 82.79 &      99.92 &      86.41 &             91.57 &      99.82 &      92.93 &                 79.79 & \textbf{99.96} &      84.70 &                 80.34 &      99.96 &      85.31 &       15.05 &      3.37 &      0.61 \\
       & \textbf{densenet201} &             90.87 &      99.77 &      92.07 &                 80.19 &      99.92 &      84.32 &                 79.93 &      99.95 &      84.38 &             91.49 &      99.87 &      93.19 &                 83.56 &      99.94 &      87.27 &                 83.79 &      99.94 &      87.38 &       20.86 &      4.07 &      0.78 \\
       & \textbf{resnext5032x4d} &             91.88 &      99.80 &      93.58 &                 80.49 &      99.92 &      84.50 &                 81.10 &      99.94 &      85.74 &             92.85 &      99.85 &      94.17 &                 82.47 &      99.95 &      86.62 &                 80.08 & \textbf{99.97} &      85.19 &       25.58 &      1.42 &      0.29 \\
       & \textbf{dpn68} &             90.73 &      99.83 &      92.78 &                 75.16 &      99.89 &      78.71 &                 79.47 &      99.93 &      84.11 &             93.93 &      99.83 &      94.75 &                 80.32 &      99.93 &      84.19 &                 83.48 &      99.93 &      87.10 &       13.92 &      2.07 &      0.42 \\
       & \textbf{dpn98} &             93.01 &      99.81 &      94.29 &                 79.04 &      99.94 &      83.64 &                 80.86 &      99.94 &      85.50 &             89.78 &      99.83 &      91.82 &                 80.72 &      99.94 &      84.82 &                 82.14 &      99.94 &      86.47 &       61.91 &      3.10 &      0.64 \\
       & \textbf{mobilenetv2} &             93.18 &      99.76 &      94.24 &                 74.74 &      99.95 &      79.60 &                 82.64 &      99.93 &      86.42 &             90.82 &      99.90 &      93.05 &                 81.65 &      99.94 &      86.07 &                 81.17 &      99.96 &      86.11 &        4.21 &      0.87 &      0.15 \\
       & \textbf{xception} &             90.82 &      99.85 &      92.51 &                 78.57 &      99.95 &      83.68 &                 81.46 & \textbf{99.96} &      86.25 &             92.97 &      99.85 &      94.36 &                 83.78 &      99.95 & \textbf{87.56} &                 82.81 &      99.93 &      86.57 &       23.24 &      0.90 &      0.16 \\
       & \textbf{inceptionv4} &             91.13 &      99.80 &      93.02 &                 80.67 &      99.94 &      84.73 &                 81.30 &      99.95 &      85.90 &             93.35 &      99.83 &      94.70 &                 80.35 & \textbf{99.96} &      85.27 &                 78.03 &      99.96 &      82.77 &       43.57 &      2.37 &      0.42 \\
       & \textbf{efficientnet-b0} &             90.76 &      99.81 &      92.17 &                 80.29 &      99.95 &      84.97 &                 79.26 & \textbf{99.96} &      84.68 &             90.17 &      99.87 &      91.92 &                 80.45 &      99.94 &      84.51 &                 82.38 &      99.94 &      86.56 &        5.76 &      1.35 &      0.25 \\
       & \textbf{efficientnet-b1} &             90.76 &      99.83 &      92.56 &                 81.40 &      99.90 &      84.32 &                 82.05 &      99.93 &      86.09 &             92.43 &      99.83 &      93.62 &                 81.60 &      99.95 &      85.96 &                 83.78 &      99.94 &      87.62 &        8.26 &      1.90 &      0.34 \\
       & \textbf{efficientnet-b2} &             91.42 &      99.85 &      93.12 &                 81.31 &      99.94 &      84.98 &                 83.12 &      99.94 &      87.15 &             90.39 &      99.88 &      92.25 &                 79.02 & \textbf{99.96} &      83.85 &                 82.82 &      99.95 &      87.29 &        9.46 &      1.92 &      0.34 \\
       & \textbf{efficientnet-b3} &             89.29 &      99.84 &      91.07 &                 82.14 &      99.93 &      85.77 &                 79.21 & \textbf{99.96} &      84.24 &             91.98 &      99.87 &      93.51 &            \textbf{84.28} &      99.91 &      86.46 &                 83.65 &      99.94 &      87.56 &       12.47 &      2.07 &      0.37 \\
       & \textbf{efficientnet-b4} &             90.25 &      99.86 &      92.28 &                 78.99 & \textbf{99.96} &      84.11 &                 81.91 &      99.92 &      85.74 &             93.08 &      99.88 &      94.71 &                 82.02 &      99.93 &      85.95 &                 85.16 &      99.94 & \textbf{89.00} &       19.35 &      2.50 &      0.46 \\
       & \textbf{efficientnet-b5} &             89.27 &      99.88 &      91.31 &                 83.52 &      99.93 &      86.70 &                 80.34 &      99.92 &      84.43 &             92.32 &      99.88 &      93.85 &                 80.82 &      99.95 &      85.46 &                 82.17 &      99.96 &      86.79 &       30.17 &      3.13 &      0.55 \\
       & \textbf{efficientnet-b6} &             89.55 &      99.83 &      91.02 &                 80.56 &      99.94 &      85.03 &                 80.67 &      99.94 &      84.93 &             92.34 &      99.88 &      94.07 &                 81.48 &      99.94 &      85.71 &                 81.13 &      99.96 &      86.03 &       42.59 &      3.51 &      0.62 \\
       & \textbf{Mean} &             90.80 &      99.81 &      92.44 &                 79.19 &      99.92 &      83.25 &                 81.15 &      99.93 &      85.46 &             91.24 &      99.85 &      92.90 &                 81.03 &      99.94 &      85.29 &                 82.27 &      99.95 &      86.52 &       23.53 &      1.88 &      0.34 \\
       & \textbf{Std} &              1.46 &       0.05 &       1.29 &                  2.49 &       0.02 &       2.31 &                  1.08 &       0.02 &       0.83 &              1.45 &       0.03 &       1.25 &                  1.59 &       0.01 &       1.17 &                  1.43 &       0.01 &       1.10 &       15.77 &      1.08 &      0.20 \\
\cline{1-23}
\multirow{27}{*}{\textbf{PSPNet}} & \textbf{vgg11} &             90.77 &      99.68 &      92.34 &                 79.44 &      99.90 &      82.51 &                 79.72 &      99.95 &      84.75 &             89.96 &      99.80 &      91.80 &                 81.31 &      99.93 &      85.41 &                 83.72 &      99.91 &      86.98 &       10.01 & \textbf{0.20} & \textbf{0.04} \\
       & \textbf{vgg13} &             89.43 &      99.74 &      91.74 &                 75.65 &      99.92 &      80.23 &                 80.85 &      99.94 &      85.44 &             89.39 &      99.78 &      91.36 &                 79.87 &      99.94 &      84.72 &                 82.64 &      99.93 &      86.64 &       10.20 &      0.22 & \textbf{0.04} \\
       & \textbf{vgg19} &             87.39 &      99.75 &      89.77 &                 72.30 &      99.92 &      76.95 &                 81.18 &      99.92 &      85.31 &             89.19 &      99.82 &      91.24 &                 79.26 &      99.95 &      83.94 &                 81.62 &      99.93 &      85.93 &       20.82 &      0.27 &      0.05 \\
       & \textbf{resnet18} &             86.38 &      99.75 &      89.21 &                 76.12 &      99.91 &      80.29 &                 79.40 &      99.93 &      83.87 &             90.05 &      99.87 &      92.50 &                 83.32 &      99.90 &      86.07 &                 83.19 &      99.91 &      86.49 &       11.32 &      0.27 &      0.05 \\
       & \textbf{resnet34} &             84.05 &      99.74 &      87.35 &                 78.26 &      99.87 &      80.64 &                 79.20 &      99.91 &      83.19 &             91.57 &      99.75 &      92.67 &                 84.05 &      99.87 &      85.12 &                 81.37 &      99.92 &      85.09 &       21.43 &      0.35 &      0.06 \\
       & \textbf{resnet50} &             89.58 &      99.79 &      91.96 &                 75.51 &      99.90 &      79.38 &                 80.29 &      99.94 &      84.83 &             92.29 &      99.76 &      93.28 &                 79.45 &      99.93 &      84.01 &                 79.33 &      99.95 &      84.49 &       24.29 &      0.48 &      0.08 \\
       & \textbf{resnet101} &             90.55 &      99.71 &      92.12 &                 77.97 &      99.87 &      80.25 &                 84.28 &      99.88 &      86.35 &             90.65 &      99.81 &      92.47 &                 77.04 &      99.94 &      81.22 &                 81.72 &      99.90 &      84.79 &       43.29 &      0.47 &      0.08 \\
       & \textbf{resnet152} &             90.25 &      99.81 &      92.47 &                 77.15 &      99.91 &      81.20 &                 84.10 &      99.89 &      86.36 &             92.50 &      99.77 &      93.75 &                 79.94 &      99.94 &      84.34 &                 81.55 &      99.92 &      85.43 &       58.93 &      0.65 &      0.11 \\
       & \textbf{densenet121} &             89.29 &      99.82 &      91.99 &                 79.47 &      99.85 &      81.30 &                 78.87 &      99.90 &      83.02 &             91.90 &      99.80 &      93.29 &                 83.83 &      99.92 &      86.29 &                 80.85 &      99.88 &      83.77 &        7.74 &      0.79 &      0.15 \\
       & \textbf{densenet161} &             89.07 &      99.82 &      91.53 &                 79.95 &      99.85 &      81.13 &                 77.58 &      99.91 &      81.97 &             91.68 &      99.80 &      93.29 &                 81.45 &      99.93 &      84.83 &                 83.78 &      99.89 &      86.25 &       27.85 &      0.82 &      0.16 \\
       & \textbf{densenet169} &             87.17 &      99.90 &      90.28 &                 80.23 &      99.90 &      83.24 &                 82.46 &      99.89 &      85.65 &             91.97 &      99.80 &      93.55 &                 80.52 &      99.94 &      85.12 &                 83.48 &      99.90 &      86.35 &       13.27 &      0.80 &      0.15 \\
       & \textbf{densenet201} &             91.04 &      99.76 &      92.85 &                 77.43 &      99.92 &      81.60 &                 80.26 &      99.89 &      83.71 &             91.32 &      99.75 &      92.35 &                 83.57 &      99.93 &      86.66 &                 81.54 &      99.93 &      85.77 &       18.88 &      0.82 &      0.16 \\
       & \textbf{resnext5032x4d} &             89.81 &      99.81 &      91.86 &                 79.45 &      99.91 &      82.77 &                 80.74 &      99.94 &      85.29 &             90.94 &      99.83 &      92.78 &                 79.69 &      99.94 &      84.05 &                 83.72 &      99.90 &      86.39 &       23.77 &      0.70 &      0.13 \\
       & \textbf{dpn68} &             81.71 &      99.81 &      85.86 &                 79.86 &      99.92 &      83.96 &                 81.38 &      99.93 &      85.33 &             91.33 &      99.78 &      92.89 &                 79.85 &      99.93 &      83.46 &                 79.29 &      99.93 &      83.93 &       12.21 &      0.79 &      0.15 \\
       & \textbf{dpn98} &             89.81 &      99.76 &      91.52 &                 79.43 &      99.93 &      83.27 &                 79.15 &      99.94 &      83.82 &             91.67 &      99.77 &      92.70 &                 80.46 &      99.94 &      84.86 &                 79.01 &      99.93 &      83.45 &       60.26 &      1.05 &      0.21 \\
       & \textbf{mobilenetv2} &             89.07 &      99.74 &      91.31 &                 80.03 &      99.88 &      82.93 &                 83.09 &      99.91 &      86.07 &             89.12 &      99.83 &      91.63 &                 75.43 & \textbf{99.96} &      81.07 &                 83.76 &      99.91 &      86.17 &   \textbf{2.26} &      0.37 &      0.06 \\
       & \textbf{xception} &             90.17 &      99.77 &      92.17 &                 77.62 &      99.91 &      81.83 &                 78.86 &      99.95 &      83.87 &             90.72 &      99.85 &      92.67 &                 81.98 &      99.83 &      83.37 &                 81.17 &      99.93 &      85.52 &       21.14 &      0.27 &      0.05 \\
       & \textbf{inceptionv4} &             90.12 &      99.74 &      91.70 &                 81.33 &      99.86 &      83.16 &                 81.99 &      99.89 &      85.58 &             90.55 &      99.78 &      91.93 &                 78.91 &      99.95 &      83.83 &                 82.03 &      99.89 &      85.20 &       41.69 &      0.64 &      0.11 \\
       & \textbf{efficientnet-b0} &             87.38 &      99.79 &      89.46 &                 82.90 &      99.91 &      85.03 &                 82.60 &      99.92 &      86.33 &             90.83 &      99.79 &      92.41 &                 80.64 &      99.93 &      84.81 &                 82.40 &      99.94 &      86.77 &        4.06 &      0.50 &      0.09 \\
       & \textbf{efficientnet-b1} &             88.12 &      99.76 &      90.23 &                 82.84 &      99.91 &      85.67 &                 80.69 &      99.94 &      85.27 &             87.93 &      99.81 &      90.19 &                 81.82 &      99.94 &      86.10 &                 81.46 &      99.93 &      85.85 &        6.56 &      0.79 &      0.14 \\
       & \textbf{efficientnet-b2} &             86.03 &      99.86 &      88.71 &                 81.00 &      99.92 &      84.34 &                 80.30 &      99.93 &      84.72 &             88.50 &      99.79 &      90.39 &                 80.84 &      99.94 &      85.30 &                 82.98 &      99.95 &      87.22 &        7.76 &      0.77 &      0.14 \\
       & \textbf{efficientnet-b3} &             88.52 &      99.80 &      90.82 &                 80.02 &      99.94 &      84.27 &                 81.07 &      99.92 &      84.96 &             88.49 &      99.84 &      90.91 &                 81.52 & \textbf{99.96} &      85.99 &                 81.39 &      99.94 &      85.73 &       10.75 &      0.78 &      0.14 \\
       & \textbf{efficientnet-b4} &             89.11 &      99.79 &      90.80 &                 81.85 &      99.91 &      85.25 &                 80.76 &      99.93 &      84.88 &             88.50 &      99.88 &      90.97 &                 81.05 &      99.92 &      85.35 &                 83.74 &      99.93 &      87.66 &       17.61 &      0.90 &      0.16 \\
       & \textbf{efficientnet-b5} &             89.09 &      99.83 &      91.10 &                 80.49 &      99.92 &      84.17 &                 83.04 &      99.90 &      86.61 &             89.26 &      99.85 &      91.54 &                 80.29 &      99.95 &      84.60 &                 83.37 &      99.94 &      87.24 &       28.42 &      1.12 &      0.20 \\
       & \textbf{efficientnet-b6} &             89.17 &      99.80 &      91.05 &                 80.14 &      99.95 &      85.00 &                 81.78 &      99.92 &      85.61 &             87.96 &      99.81 &      89.81 &                 81.47 &      99.92 &      84.81 &                 82.87 &      99.95 &      87.34 &       40.82 &      1.24 &      0.23 \\
       & \textbf{Mean} &             88.52 &      99.78 &      90.81 &                 79.06 &      99.90 &      82.42 &                 80.95 &      99.92 &      84.91 &             90.33 &      99.80 &      92.09 &                 80.70 &      99.93 &      84.61 &                 82.08 &      99.92 &      85.86 &       21.81 &      0.64 &      0.12 \\
       & \textbf{Std} &              2.13 &       0.05 &       1.63 &                  2.39 &       0.03 &       2.08 &                  1.66 &       0.02 &       1.15 &              1.39 &       0.03 &       1.06 &                  1.93 &       0.03 &       1.33 &                  1.41 &       0.02 &       1.12 &       15.73 &      0.29 &      0.05 \\
\cline{1-23}
\multirow{2}{*}{\textbf{Global}} & \textbf{Mean} &             90.64 &      99.81 &      92.30 &                 80.36 &      99.91 &      83.56 &                 81.41 &      99.93 &      85.44 &             91.61 &      99.85 &      93.18 &                 81.53 &      99.94 &      85.47 &                 82.51 &      99.94 &      86.44 &       24.64 &      1.60 &      0.29 \\
       & \textbf{Std} &              2.36 &       0.06 &       1.82 &                  2.32 &       0.02 &       1.93 &                  1.44 &       0.02 &       1.08 &              1.60 &       0.04 &       1.30 &                  1.59 &       0.02 &       1.17 &                  1.32 &       0.02 &       1.04 &       16.99 &      1.11 &      0.21 \\
\bottomrule
\end{tabular}

	\end{adjustbox}
\end{table*}

\subsection{Architecture comparison}
\begin{figure}[!t]
	\centering
	\rotatebox[origin=l]{90}{\scriptsize Lung segmentation}\subfloat{\includegraphics[width=0.24\textwidth]{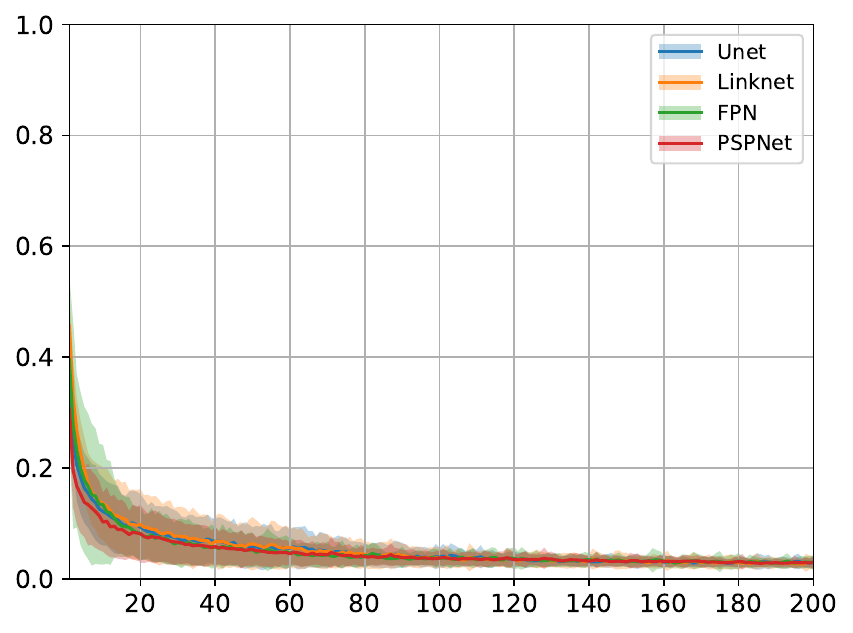}}
	\subfloat{\includegraphics[width=0.24\textwidth]{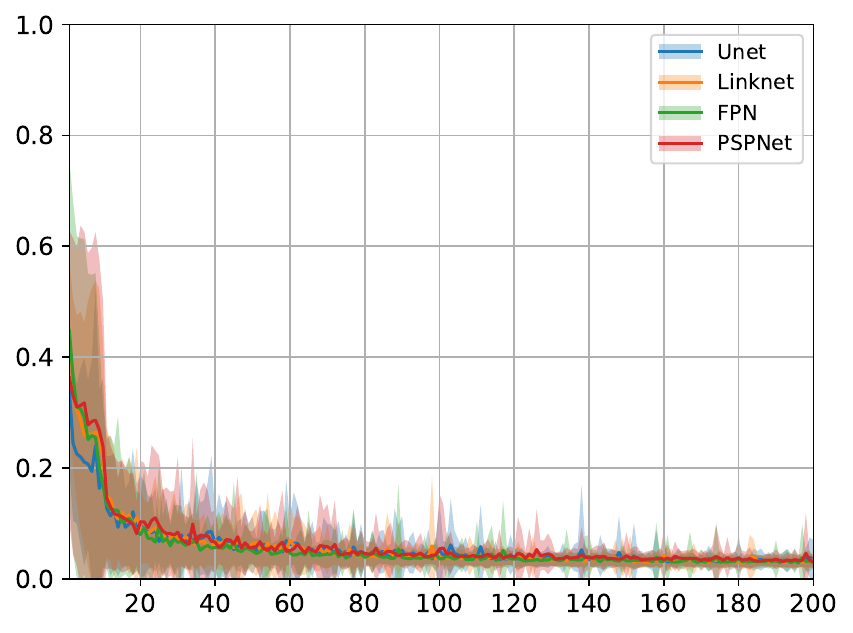}}
	\\
	\rotatebox[origin=l]{90}{\scriptsize Lesion segmentation A}\subfloat{\includegraphics[width=0.24\textwidth]{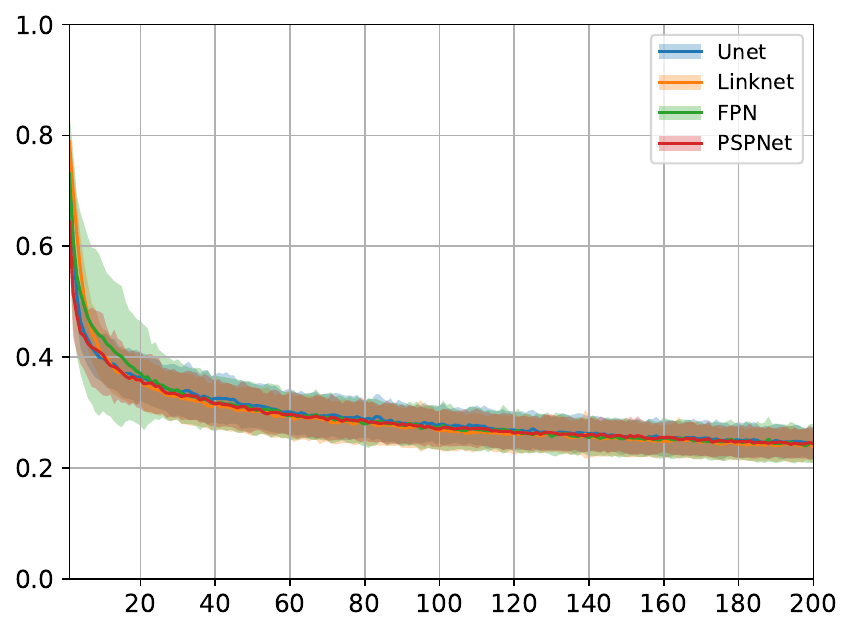}}
	\subfloat{\includegraphics[width=0.24\textwidth]{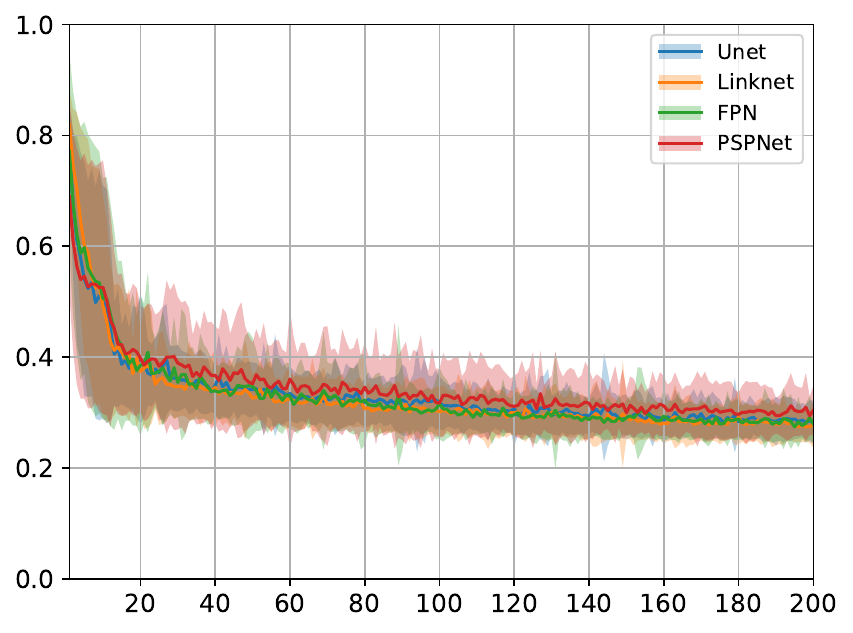}}
	\\
	\rotatebox[origin=l]{90}{\scriptsize Lesion segmentation B}\subfloat{\includegraphics[width=0.24\textwidth]{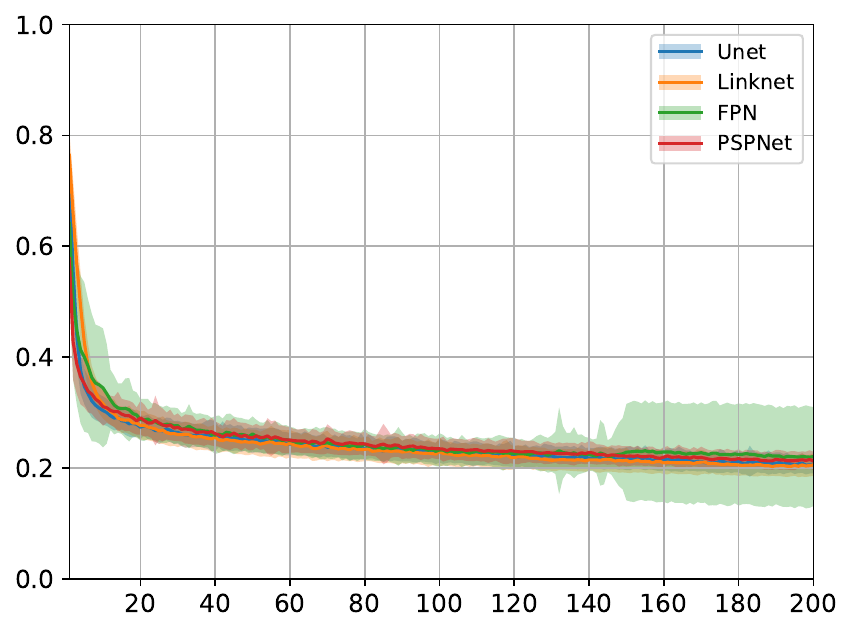}}
	\subfloat{\includegraphics[width=0.24\textwidth]{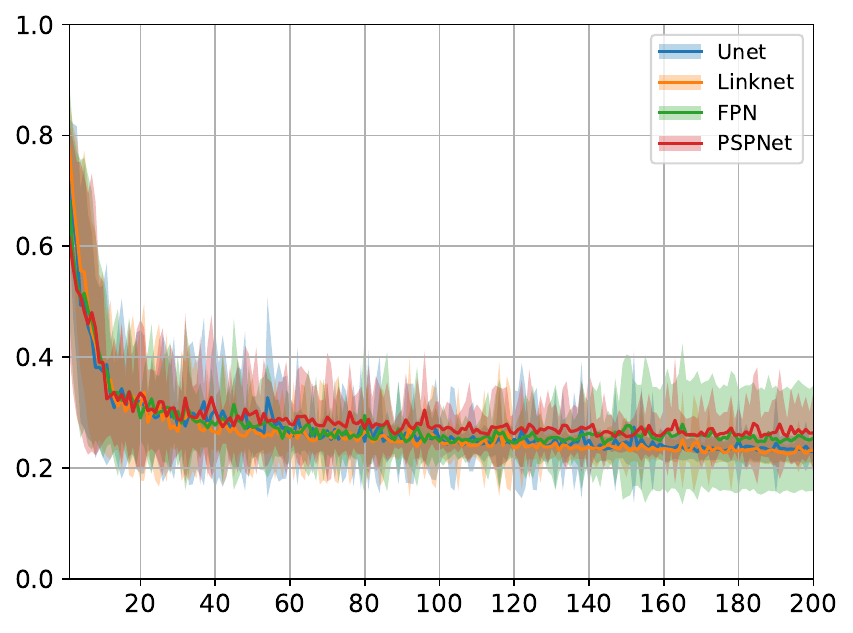}}
	\\
	\setcounter{subfigure}{0}
	\rotatebox[origin=l]{90}{\scriptsize Lesion segmentation B-A}\subfloat[Train loss]{\includegraphics[width=0.24\textwidth]{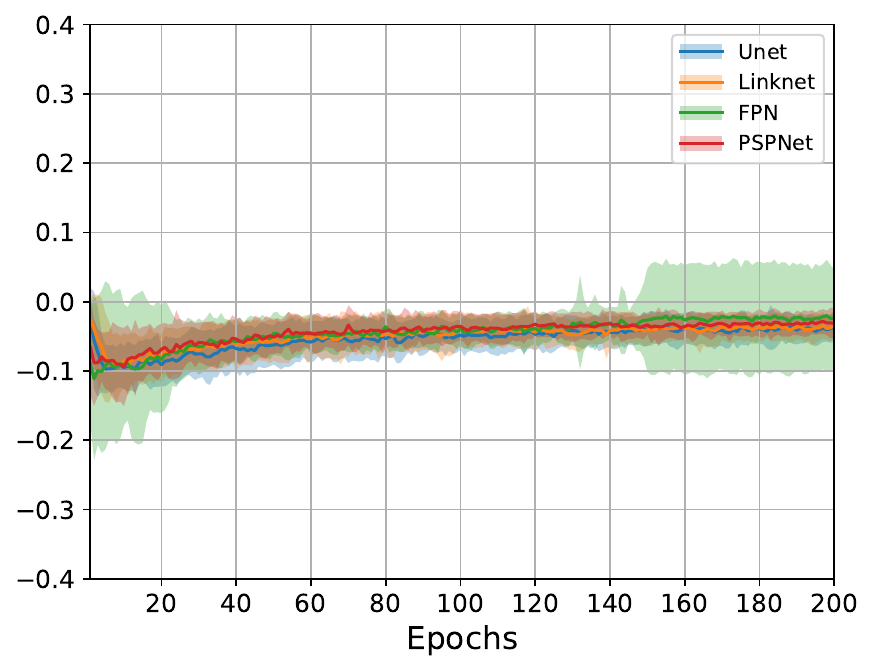}}
	\subfloat[Validation loss]{\includegraphics[width=0.24\textwidth]{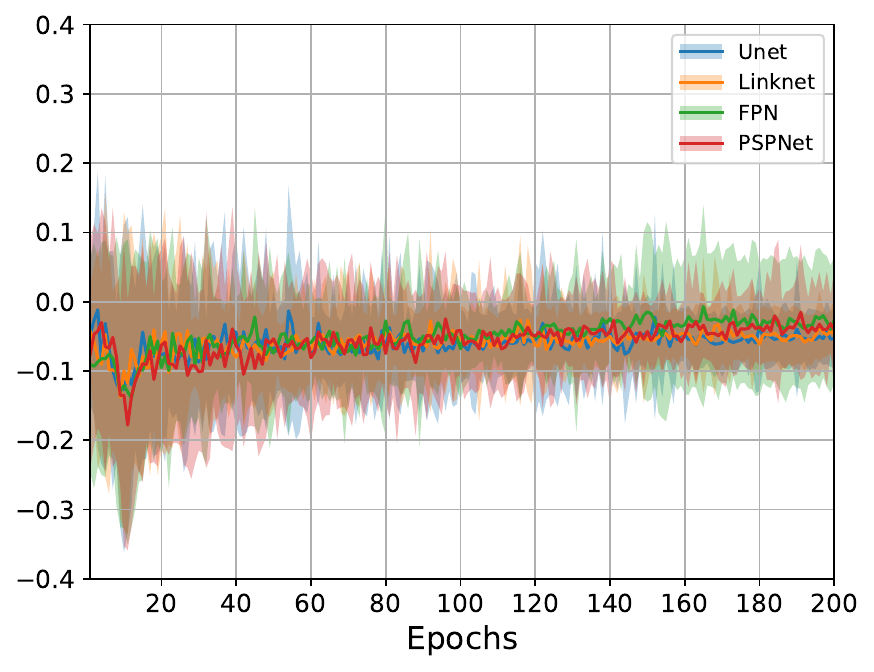}}
	\caption{Training and validation loss plots with mean (line) and standard deviation (patch) for the four architectures segregated upon encoders for each experimental setup.
	Rows correspond to experimental setups and the difference between lesion segmentation B and A and columns correspond to train and validation losses.}\label{fig:loss}
\end{figure}

\begin{figure}[!t]
	\centering
	\rotatebox[origin=l]{90}{\hspace{1.5em}\scriptsize Dice (\%)}\subfloat[Lung segmentation]{\includegraphics[width=0.24\textwidth]{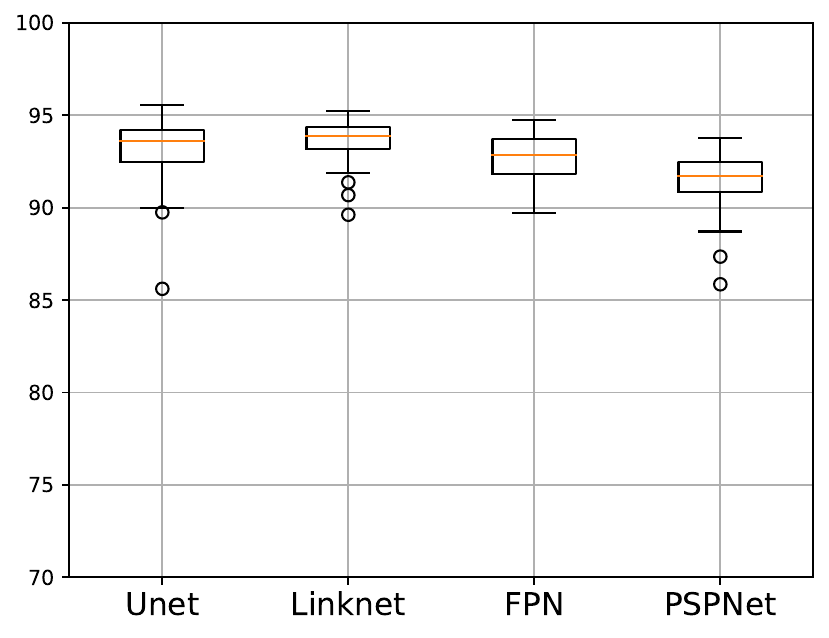}}
	\subfloat[Lesion segmentation A]{\includegraphics[width=0.24\textwidth]{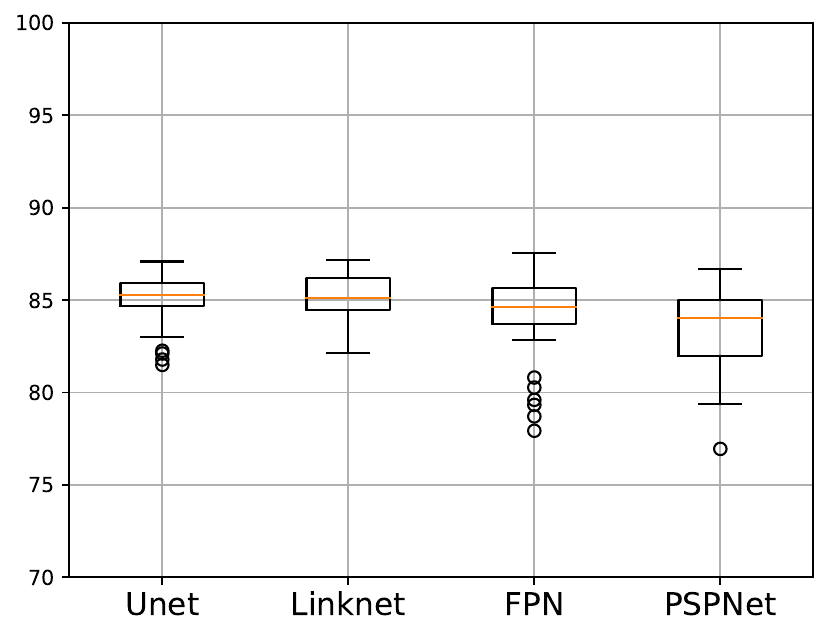}}
	\\
	\rotatebox[origin=l]{90}{\hspace{1.5em}\scriptsize Dice (\%)}\subfloat[Lesion segmentation B]{\includegraphics[width=0.24\textwidth]{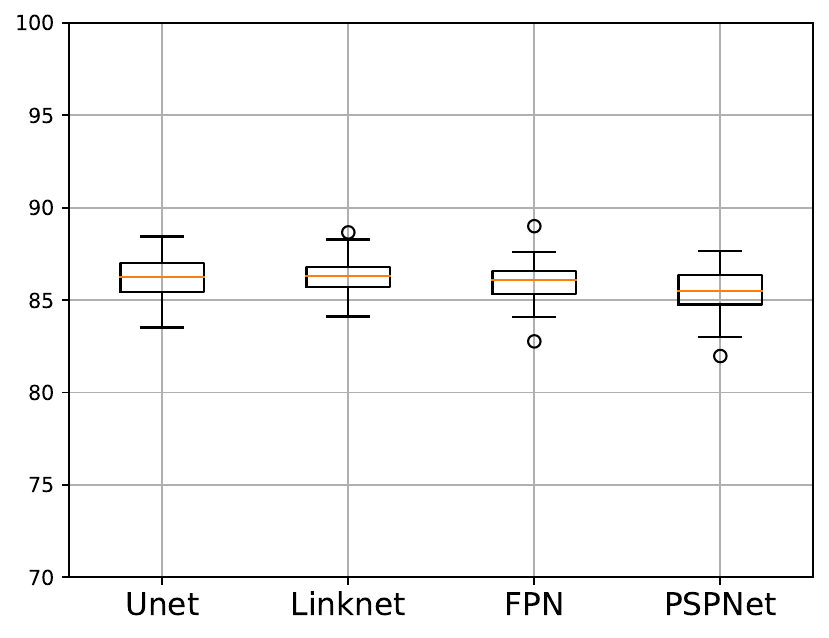}}
	\subfloat[Initialization]{\includegraphics[width=0.24\textwidth]{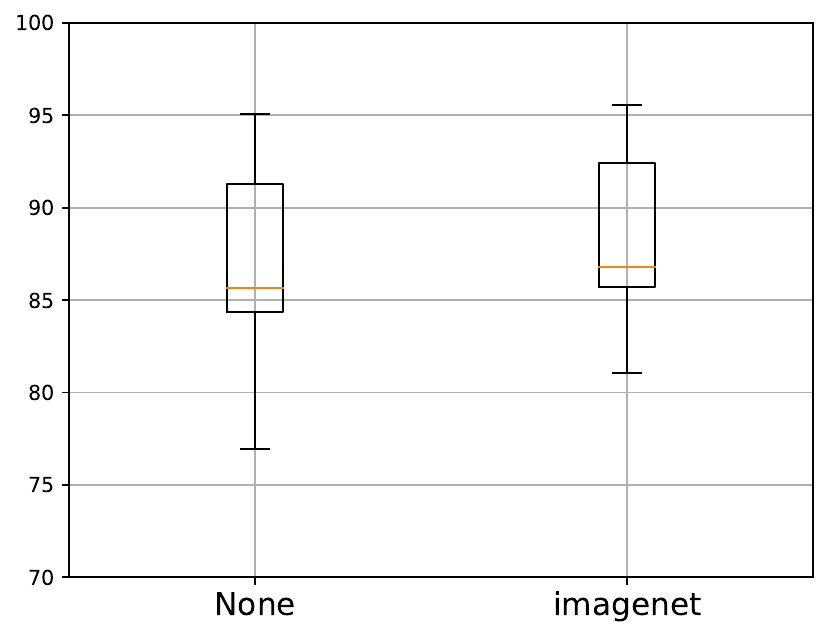}}
	\caption{Boxplots of Dice for the three experiments and for each architecture in (a), (b) and (c) and weight initialization for all experiments in (d).}\label{fig:box}
\end{figure}

In Fig.~\ref{fig:loss} the train and validation loss vs.\ epochs of the four architectures segregated upon encoders is depicted.
In all experiments and architectures, the training loss during the $5$ first epochs, decreases fast and in a slower rate during the next epochs.
We can observe the same behaviour for validation loss during the $15$ first epochs but with more variability, which can be explained by the use of the dice loss as a validation metric.
More specifically, we observe faster convergence for PSPNet for training loss compared to the other architectures, greater variance for FPN and lower convergence for Linknet in both training and validation.
In Fig.~\ref{fig:box} the Dice boxplots for the three experiments is plotted.
Regarding time performance for training and inference, the fastest architecture is PSPNet and the slowest is Linknet even having more parameters than Unet.

The mean Dice results for Unet are $\DTLfetch{keys-values}{key}{lung-segmentation-Unet-imagenet-mean}{value}\%\pm\DTLfetch{keys-values}{key}{lung-segmentation-Unet-imagenet-std}{value}\%$ for lung segmentation, $\DTLfetch{keys-values}{key}{lesion-segmentation-a-Unet-imagenet-mean}{value}\%\pm\DTLfetch{keys-values}{key}{lesion-segmentation-a-Unet-imagenet-std}{value}\%$ for `lesion segmentation A', $\DTLfetch{keys-values}{key}{lesion-segmentation-b-Unet-imagenet-mean}{value}\%\pm\DTLfetch{keys-values}{key}{lesion-segmentation-b-Unet-imagenet-std}{value}\%$ for `lesion segmentation B'.
The mean Dice results for Linknet are $\DTLfetch{keys-values}{key}{lung-segmentation-Linknet-imagenet-mean}{value}\%\pm\DTLfetch{keys-values}{key}{lung-segmentation-Linknet-imagenet-std}{value}\%$ for lung segmentation, $\DTLfetch{keys-values}{key}{lesion-segmentation-a-Linknet-imagenet-mean}{value}\%\pm\DTLfetch{keys-values}{key}{lesion-segmentation-a-Linknet-imagenet-std}{value}\%$ for `lesion segmentation A', $\DTLfetch{keys-values}{key}{lesion-segmentation-b-Linknet-imagenet-mean}{value}\%\pm\DTLfetch{keys-values}{key}{lesion-segmentation-b-Linknet-imagenet-std}{value}\%$ for `lesion segmentation B'.
The mean Dice results for FPN are $\DTLfetch{keys-values}{key}{lung-segmentation-FPN-imagenet-mean}{value}\%\pm\DTLfetch{keys-values}{key}{lung-segmentation-FPN-imagenet-std}{value}\%$ for lung segmentation, $\DTLfetch{keys-values}{key}{lesion-segmentation-a-FPN-imagenet-mean}{value}\%\pm\DTLfetch{keys-values}{key}{lesion-segmentation-a-FPN-imagenet-std}{value}\%$ for `lesion segmentation A', $\DTLfetch{keys-values}{key}{lesion-segmentation-b-FPN-imagenet-mean}{value}\%\pm\DTLfetch{keys-values}{key}{lesion-segmentation-b-FPN-imagenet-std}{value}\%$ for `lesion segmentation B'.
The mean Dice results for PSPNet are $\DTLfetch{keys-values}{key}{lung-segmentation-PSPNet-imagenet-mean}{value}\%\pm\DTLfetch{keys-values}{key}{lung-segmentation-PSPNet-imagenet-std}{value}\%$ for lung segmentation, $\DTLfetch{keys-values}{key}{lesion-segmentation-a-PSPNet-imagenet-mean}{value}\%\pm\DTLfetch{keys-values}{key}{lesion-segmentation-a-PSPNet-imagenet-std}{value}\%$ for `lesion segmentation A', $\DTLfetch{keys-values}{key}{lesion-segmentation-b-PSPNet-imagenet-mean}{value}\%\pm\DTLfetch{keys-values}{key}{lesion-segmentation-b-PSPNet-imagenet-std}{value}\%$ for `lesion segmentation B'.

\begin{figure*}[!t]
	\rotatebox[origin=l]{90}{\hspace{1em}\scriptsize Lung segmentation}\subfloat{\includegraphics[width=0.20\textwidth]{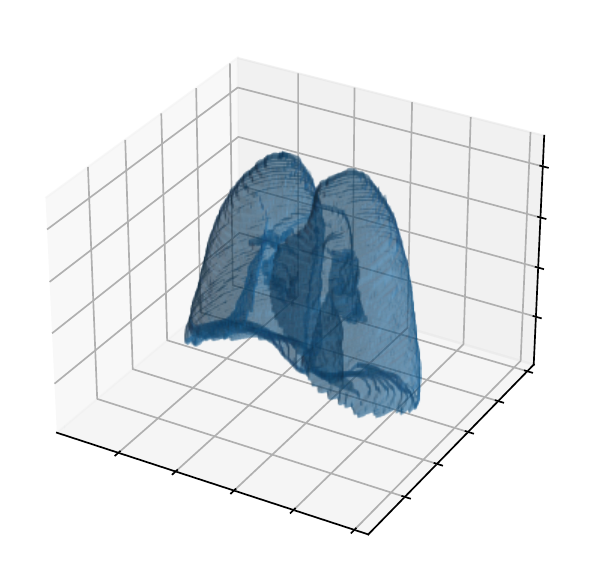}}
	\subfloat{\includegraphics[width=0.20\textwidth]{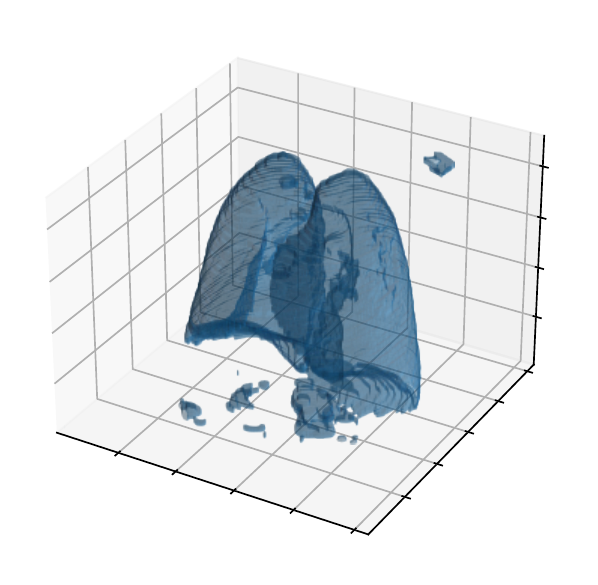}}
	\subfloat{\includegraphics[width=0.20\textwidth]{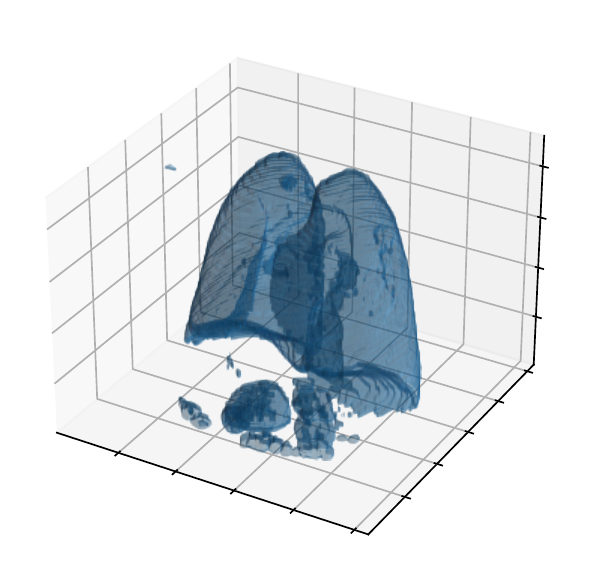}}
	\subfloat{\includegraphics[width=0.20\textwidth]{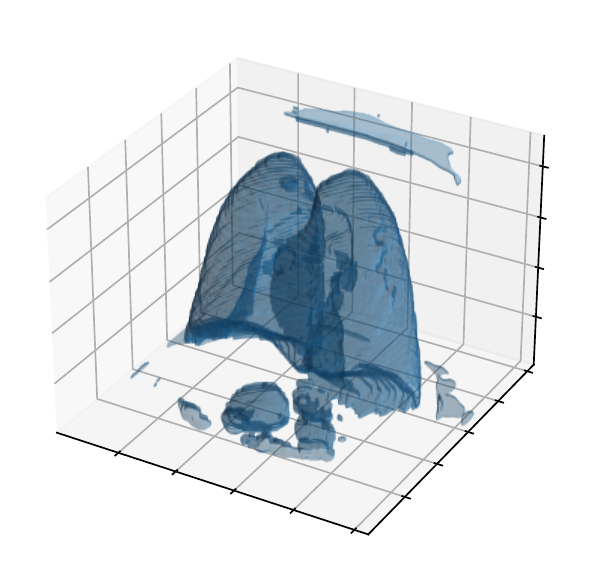}}
	\subfloat{\includegraphics[width=0.20\textwidth]{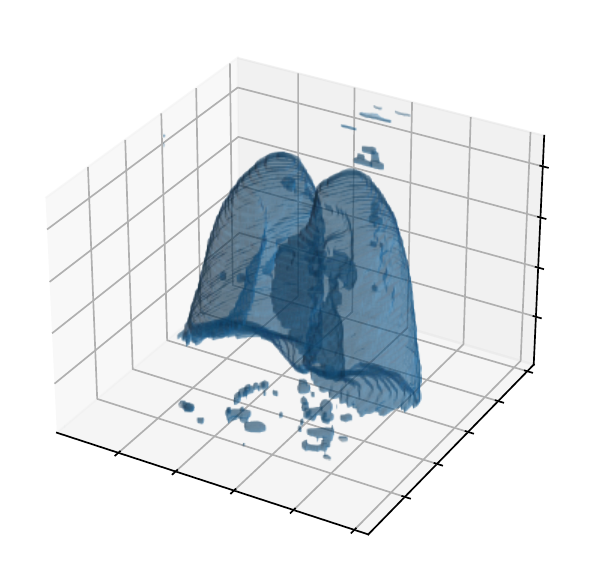}}
	\\
	\rotatebox[origin=l]{90}{\hspace{0.8em}\scriptsize Lesion segmentation A}\subfloat{\includegraphics[width=0.20\textwidth]{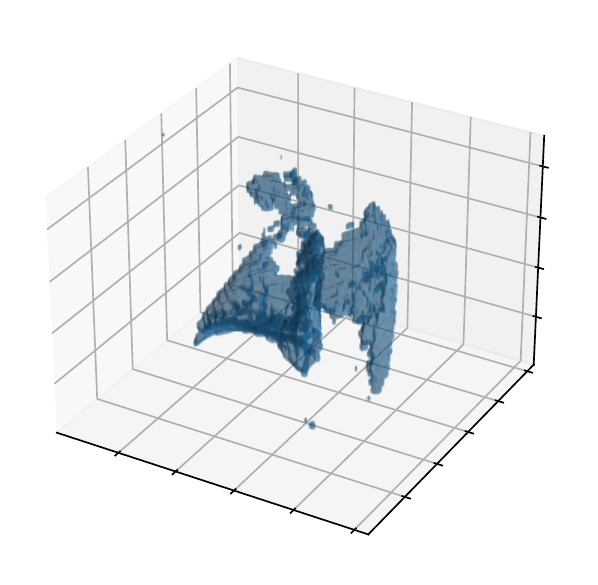}}
	\subfloat{\includegraphics[width=0.20\textwidth]{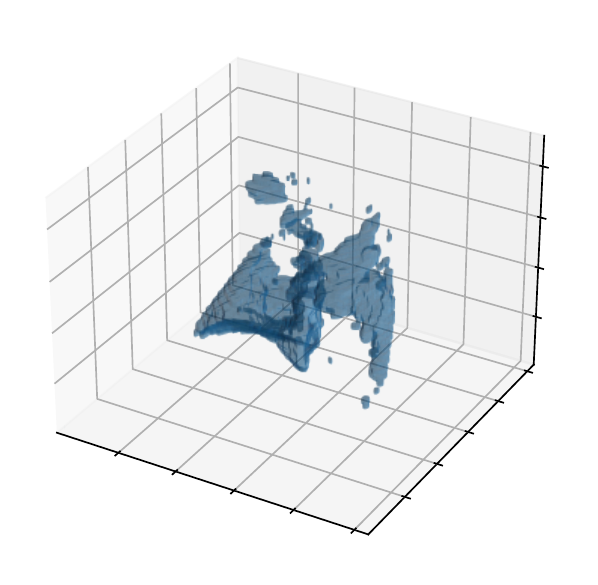}}
	\subfloat{\includegraphics[width=0.20\textwidth]{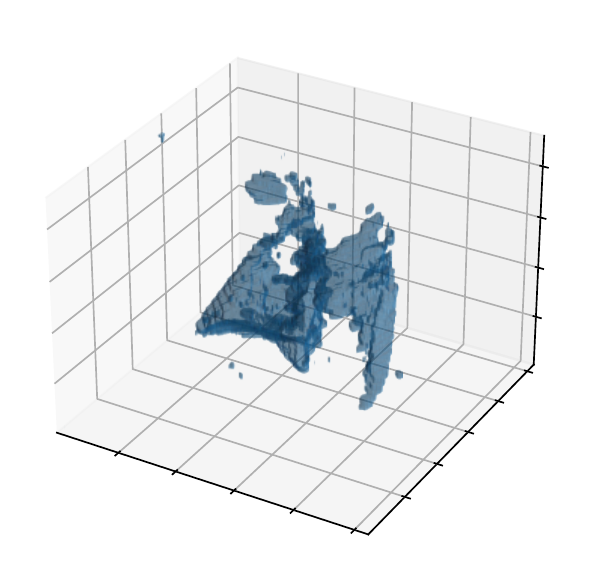}}
	\subfloat{\includegraphics[width=0.20\textwidth]{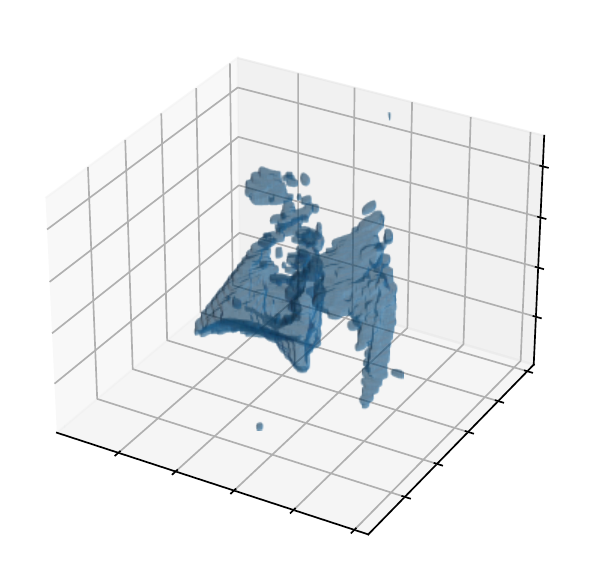}}
	\subfloat{\includegraphics[width=0.20\textwidth]{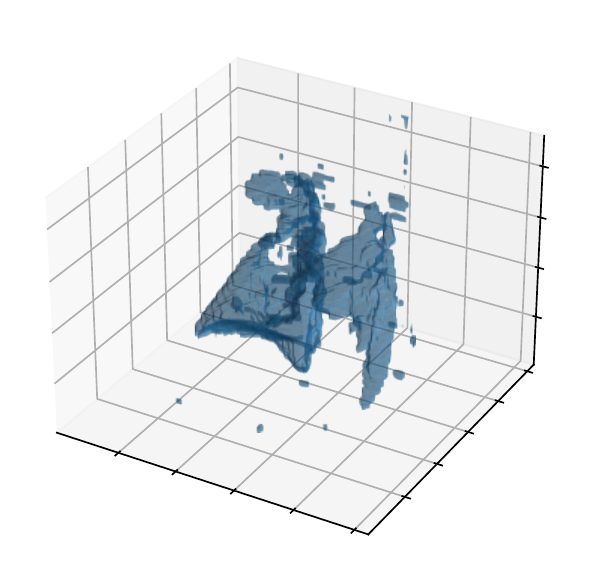}}
	\\
	\setcounter{subfigure}{0}
	\rotatebox[origin=l]{90}{\hspace{0.8em}\scriptsize Lesion segmentation B}\subfloat[Mask]{\includegraphics[width=0.20\textwidth]{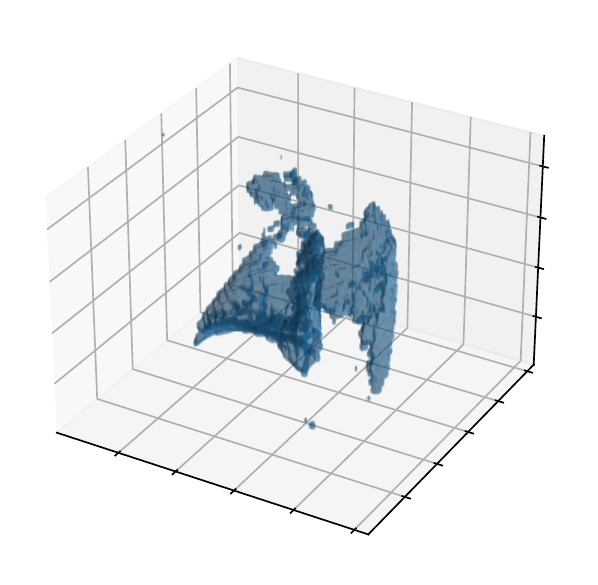}}
	\subfloat[Unet]{\includegraphics[width=0.20\textwidth]{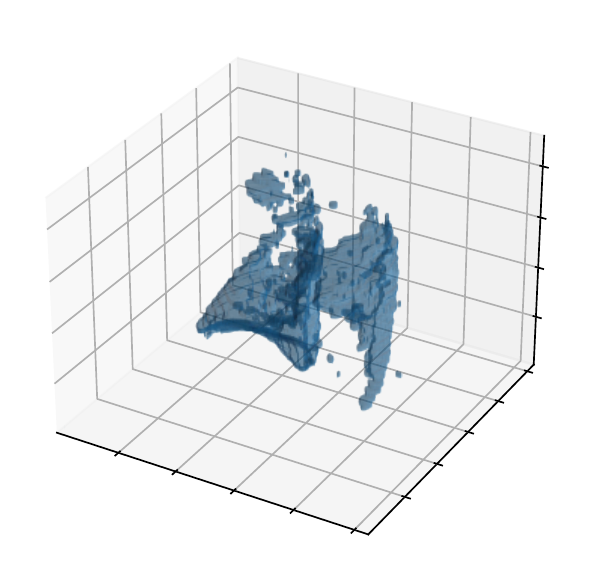}}
	\subfloat[Linknet]{\includegraphics[width=0.20\textwidth]{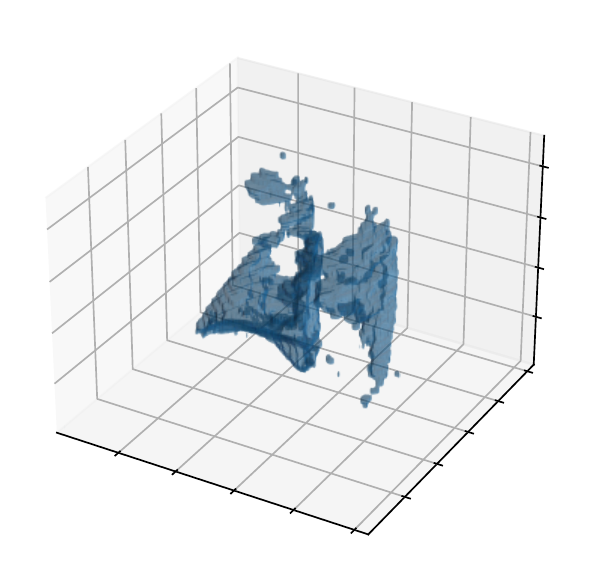}}
	\subfloat[FPN]{\includegraphics[width=0.20\textwidth]{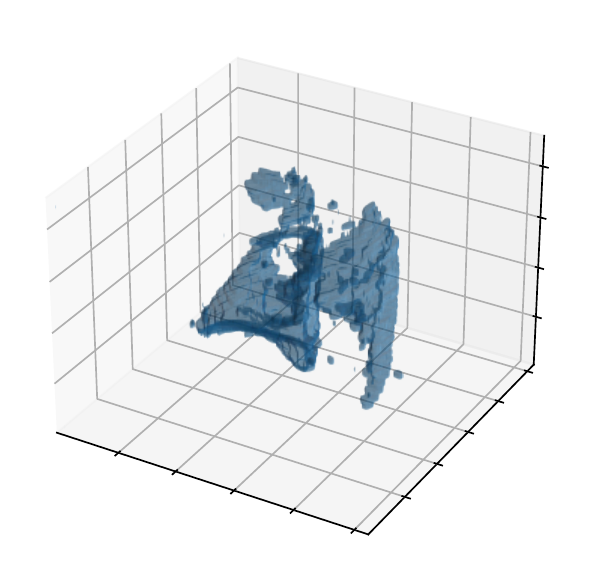}}
	\subfloat[PSPNet]{\includegraphics[width=0.20\textwidth]{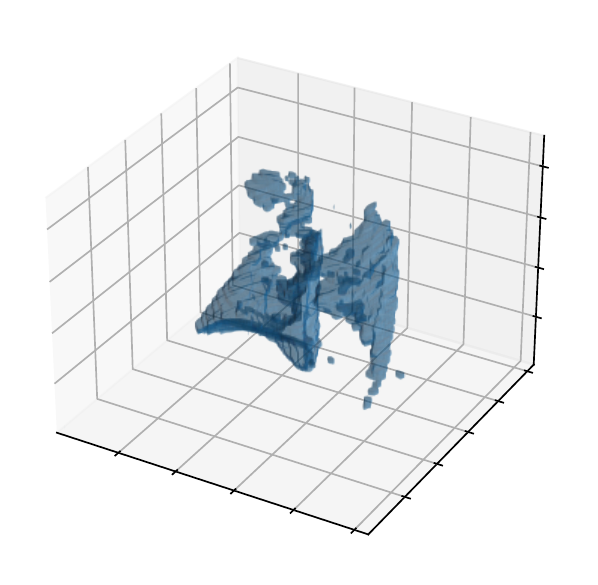}}
	\caption{Reconstructed volumes from predicted masks.
	Reconstruction was done by stacking the predicted masks on the z-axis and then applying the marching cubes algorithm.}\label{fig:volumes}
\end{figure*}

In Fig.~\ref{fig:volumes} the predictions as a volume for the three experiments, for the resnet18 encoder are visualized, demonstrating good match with the original masks.

\subsection{Lesion segmentation A vs.\ Lesion segmentation B}
In the bottom two subfigures in Fig.~\ref{fig:loss} the difference of the train and validation loss between the `lesion segmentation A' and `lesion segmentation B' are depicted.
We observe convergence between the losses, which is an indication that when training for large number of epochs the use of lung masks, as a preprocessing step, is less required.

\subsection{Random initialization vs.\ pretrained on ImageNet}
\begin{figure}[!t]
	\centering
	\rotatebox[origin=l]{90}{\scriptsize Random initialization}\subfloat{\includegraphics[width=0.24\textwidth]{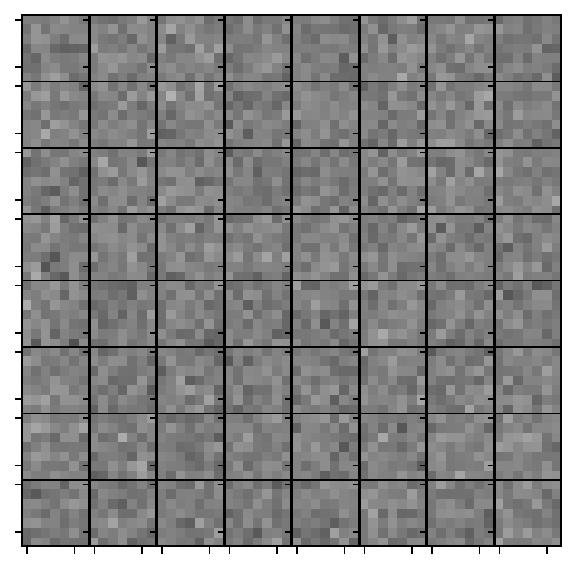}}
	\subfloat{\includegraphics[width=0.24\textwidth]{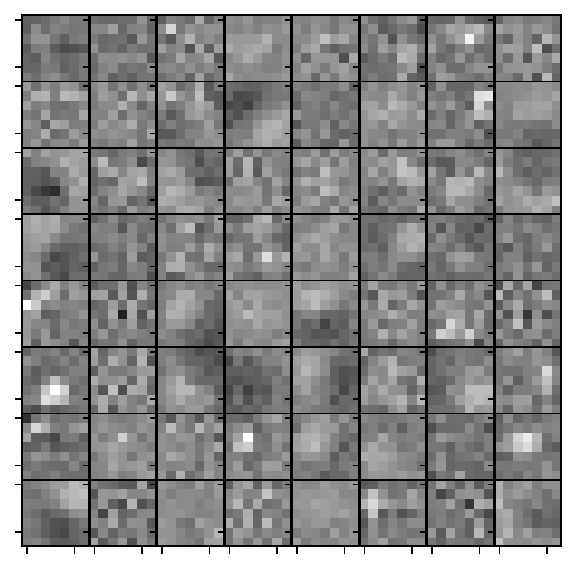}}
	\\
	\setcounter{subfigure}{0}
	\rotatebox[origin=l]{90}{\scriptsize ImageNet initialization}\subfloat[Before training]{\includegraphics[width=0.24\textwidth]{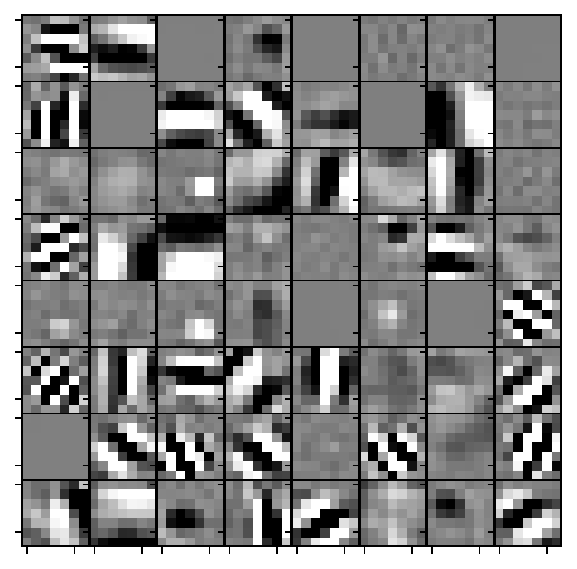}}
	\subfloat[After training]{\includegraphics[width=0.24\textwidth]{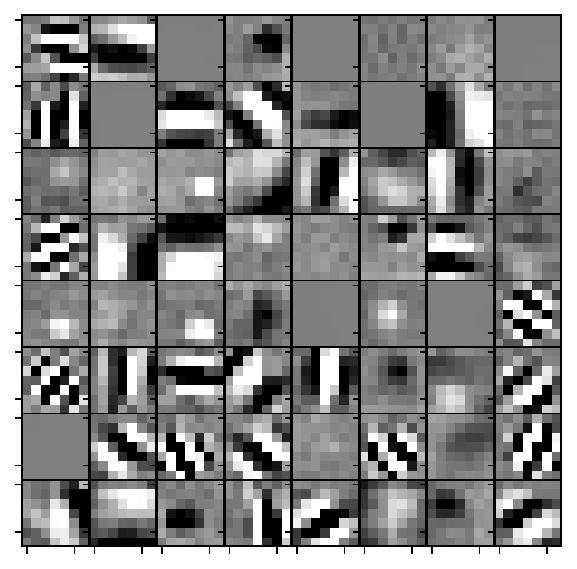}}
	\caption{Grid visualization of the $64$ weights with size $7\times 7$ of the first convolutional layer for the Unet architecture with ResNet-18 encoder.
	The two left and two right grids correspond to random and ImageNet pretrained initialization, respectively.
	The data range of the colormap of the images has been adjusted in the range $[-0.4, 0.4]$.}\label{fig:weights}
\end{figure}

In Fig.~\ref{fig:weights} the weights are depicted, in which we observe that with random initialization the weights depict high and low frequency textures after training.
The Dice for using random initialization is $\DTLfetch{keys-values}{key}{None-mean}{value}\%\pm\DTLfetch{keys-values}{key}{None-std}{value}\%$ and for ImageNet initialization is $\DTLfetch{keys-values}{key}{imagenet-mean}{value}\%\pm\DTLfetch{keys-values}{key}{imagenet-std}{value}\%$.

\section{Discussion}\label{sec:discussion}
A motivation for this study is that the large amount of new models that are proposed, rarely conduct ablation studies and do not compare with simple baselines.
This study can be used as a set of baseline models that DL model designers will test on, to confirm and evaluate whether their novel model performs better than other models, e.g.\ by comparing their accuracy with models with the same number of parameters and/or training/validation time.

A common preprocessing step for lesion segmentation is using lung masks as either from manual annotation or from an automatic method.
This step naturally improves lesion segmentation since the model only needs to search within the lung region instead of the whole image.
Moreover, the use of this step is necessary in cases where the lesion is orders of magnitude smaller than the lungs and the background, justifying the characterization of lesion datasets as `unbalanced'.
The arguments against using this preprocessing step is that the complexity of the model and the cost of annotation by the experts are increased.
The question to be answered by the expert is whether it is beneficial to increase the model complexity and the annotation cost to achieve the additional lesion segmentation accuracy increase.
Related previous work was conducted by Shi et al.~\cite{shi2020review}, which categorized COVID-19 segmentation models between:
\begin{itemize}
	\item the lung-lesion-oriented models, which directly segment lesions and
	\item the lung-region-oriented models, which first segment the lungs and then pass the masked region for further segmentation or classification.
\end{itemize}

Regarding the encoder weight initialization experiment we confirm previous research such as~\cite{orsic2019defense} that pretrained weights significantly improve segmentation results, however we hypothesize that for as the number of epochs increases, the accuracy gap between them decreases.
Similar positive findings regarding transfer learning such as improved performance and faster convergence were also reported in~\cite{wang2020does}.

Limitations of this study include the use of small number of training data, however this is partially solved by the use of data augmentation methods that are applied on each training epoch.
Moreover, it is costly to gather and annotate medical images especially when extreme events such as the COVID-19 outbreak occur.
Therefore, the use of this study training dataset is representative of medical datasets that exist in the wild as summarized in~\cite{hesamian2019deep}, which contains samples in the order of $10^2$ to lower $10^3$.

Future work includes the use of neuron and layer attribution methods to investigate reasons that specific combinations of architectures and encoders perform better than others.

\section{Conclusions}\label{sec:conclusions}
The need for fast, accurate and automatic diagnosis of COVID-19 requires highly reliable and publicly available models.
We demonstrate specific properties that increase model segmentation and help experts in improved diagnosis by publicly providing pretrained models ready to be used for fine-tuning in experimental setups without GPU\@.

\section*{ACKNOWLEDGMENT}
This work received funding from the European Union’s Horizon 2020 research and innovation programme under grant agreement No 875325 (TeNDER, affecTive basEd iNtegrateD carE for betteR Quality of Life).

\bibliographystyle{elsarticle-num}
\bibliography{ms.bib}

\end{document}